\documentclass[fleqn,usenatbib]{mnras}

\usepackage{newtxtext,newtxmath}

\usepackage[T1]{fontenc}
\usepackage{ae,aecompl}

\usepackage{graphicx}	
\usepackage{amsmath}	
\usepackage{amssymb}	

\newcommand{\Msun}{\mbox{$\mathrm{M}_{\odot}$}}
\newcommand{\Rsun}{\mbox{$\mathrm{R}_{\odot}$}}

\usepackage[usenames,dvipsnames]{xcolor}
\usepackage{soul}

\newcommand{\bs}[1]{\textit{\textcolor{orange}{#1}}}

\title[Multi-wavelength observations of GD\,394]{Multi-wavelength observations of the EUV variable metal-rich white dwarf GD\,394}

\author[Wilson et al.]{David J. Wilson$^{1,2}$\thanks{djwilson394@gmail.com},
  Boris T. G{\"a}nsicke$^{1,3}$, Detlev Koester$^4$, Odette Toloza$^1$,\newauthor Jay B. Holberg$^5$, Simon P. Preval$^{6}$, Martin A. Barstow$^6$, Claudia Belardi$^6$,\newauthor Matthew R. Burleigh$^6$, Sarah L. Casewell$^6$,
  P. Wilson Cauley$^7$, Paul Chote$^1$,\newauthor  Jay Farihi$^8$, Mark A. Hollands$^1$,
  Knox S. Long$^9$, Seth Redfield$^{10}$ \medskip\\
$^{1}$ Department of Physics, University of Warwick, Coventry CV4 7AL,
UK \\
$^{2}$ McDonald Observatory, University of Texas at Austin, 2515 Speedway, C1402, Austin, TX 78712, USA  \\
$^{3}$ Centre for Exoplanets and Habitability, University of Warwick, Coventry CV4 7AL, UK \\
$^{4}$ Institut f\"ur Theoretische Physik und Astrophysik, University of Kiel,
24098 Kiel, Germany\\
$^{5}$ University of Arizona, Lunar and Planetary Lab., Tucson, Arizona, 85718, USA \\
$^{6}$ Department of Physics and Astronomy, University of Leicester, University Road, Leicester LE1 7RH, UK\\
$^{7}$ School of Earth and Space Exploration, Arizona State University, Tempe, AZ 85281, USA\\
$^{8}$ University College London, Department of Physics \& Astronomy, Gower Street, London WC1E 6BT, UK\\
$^{9}$ Space Telescope Science Institute, Baltimore, MD 21218, USA\\
$^{10}$ Wesleyan University, Department of Astronomy and Van Vleck Observatory, 96 Foss Hill Dr., Middletown, CT 06459, USA\\
}

\date{Accepted 2018 November 20. Received 2018 November 20; in original form 2018 January 15}

\pubyear{2018}

\begin{document}
\label{firstpage}
\pagerange{\pageref{firstpage}--\pageref{lastpage}}
\maketitle
\begin{abstract}
We present new {\em Hubble Space Telescope} ({\em HST}) ultraviolet and ground-based optical observations of the  hot, metal-rich white dwarf GD\,394. Extreme-ultraviolet (EUV) observations in 1992--1996 revealed a 1.15\,d periodicity with a 25\,per\,cent amplitude, hypothesised to be due to metals in a surface accretion spot. We obtained phase-resolved {\em HST}/Space Telescope Imaging Spectrograph (STIS) high-resolution far-ultraviolet (FUV) spectra of GD\,394 that sample the entire period, along with a large body of supplementary data. We find no evidence for an accretion spot, with the flux, accretion rate and radial velocity of GD\,394 constant over the observed timescales at ultraviolet and optical wavelengths. We speculate that the spot may have no longer been present when our observations were obtained, or that the EUV variability is being caused by an otherwise undetected evaporating planet. The atmospheric parameters obtained from separate fits to optical and ultraviolet spectra are inconsistent, as is found for multiple hot white dwarfs. We also detect non-photospheric, high-excitation absorption lines of multiple volatile elements, which could be evidence for a hot plasma cocoon surrounding the white dwarf.
\end{abstract}




\section{Introduction}\label{sec:intro}

\noindent Of the hundreds of known remnant planetary systems at white dwarfs, short time-scale variability has been observed at only a handful. Along with the transits at WD\,1145+017 \citep{vanderburgetal15-1, gaensickeetal16-1, redfieldetal17-1}, examples include changes in infrared flux from multiple dusty debris discs \citep{xu+jura14-1,xuetal18-1,farihietal18-1}, the growth and subsequent disappearance of gaseous emission from SDSS\,J1617+1620 over an eight year period \citep{wilsonetal14-1} and the year-to-year changes in gaseous emission line profiles at several other white dwarfs \citep{wilsonetal15-1,manseretal16-1,manseretal16-2,dennihyetal18-1}.

The first\footnotemark metal-rich, single white dwarf observed to be variable was the hot and bright ($V_{\mathrm{mag}}=13.09$) white dwarf GD\,394 (WD\,2111+489). \citet{chayeretal00-1} and \citet{vennesetal06-1} measured near-Solar abundances of Fe in the hydrogen atmosphere of GD\,394, along with high Si and P abundances. White dwarfs with $T_{\mathrm{eff}}\gtrsim$20000\,K may retain some metals in their atmospheres via radiative levitation, where outward-directed photons transfer momentum to metals and counteract the effects of downward diffusion \citep{chayeretal10-1,chayer14-1}. However, radiative levitation is predicted to support only small amounts of Fe \citep{chayeretal95-1,schuhetal02-1}, and cannot explain the Fe abundances of GD\,394 \citep{dupuisetal00-1}. Although the radiative support for Si is stronger, \citet{barstowetal96-1} showed that the Si abundance of GD\,394 similarly exceeds the predictions. The timescales for metals to diffuse out of the photospheres of hot white dwarfs are of order days \citep{paquetteetal86-1,koester+wilken06-1,koester09-1}, so for the metal abundances to be higher than those supported by radiative levitation GD\,394 must be currently, and continuously, accreting material from an external source.

\footnotetext{With the exception of pulsating objects, where the variability is inherent to the white dwarf rather than produced by external material.}

Most remarkably, \citet{dupuisetal00-1} detected a 1.15\,d periodic modulation in the extreme-ultraviolet (EUV) flux of GD\,394, with an amplitude of 25\,per\,cent. The variability was detected in three separate instruments onboard the {\em Extreme Ultraviolet Explorer} ({\em EUVE}) in observations spanning 1992--1996, leading \citet{dupuisetal00-1} to conclude that it was intrinsic to the star. Their preferred explanation for this variability was that the accreting material is being channelled, presumably by a magnetic field, onto a spot, which is rotating in and out of view over the white dwarf rotation period. The change in Si concentration within the spot would affect the atmospheric opacity in the EUV, producing the observed variability. This is similar to the explanation for the soft X-ray variable V471\,Tau, where a fraction of the wind from a K\,dwarf is magnetically funnelled onto a white dwarf companion \citep{jensenetal86-1,clemensetal92-1}. 

The observations of GD\,394 published so far mostly took place before the development of the research field of remnant planetary systems at white dwarfs, and as such planetesimal debris was not considered as a possible source for the detected metals. Early radial velocity measurements suggested that the metals were in a cloud around the white dwarf \citep{shipmanetal95-2}, but later, more precise radial velocity measurements were consistent with a photospheric origin of the metals \citep{barstowetal96-1, bannisteretal03-1}. \citet{dupuisetal00-1} favoured accretion either from the Interstellar Medium (ISM), pointing out that there is a high-density ISM clump nearby \citep{sfeiretal99-1}, or from an undetected companion. Both of these sources can be ruled out: accretion from either source would have a large mass fraction of carbon, which is not detected in the photosphere of GD\,394 \citep{dickinsonetal12-2}, and no evidence for a stellar mass companion has been observed either from radial velocity measurements \citep{safferetal98-1} or via searches for an infrared excess \citep{mullallyetal07-1}. As accretion of planetary debris is now thought to be responsible for most, if not all, metal pollution in single white dwarfs \citep{farihietal10-2,barstowetal14-1}, it is likely to also be the source of the metals at GD\,394. 

Here we test the hypothesis that the EUV variation is caused by an accretion spot in two ways: Firstly, phase-resolved spectroscopy should show changes in metal line strength as the spot moves in and out of view; secondly, flux redistribution from the spot should manifest as optical variability in anti-phase to the EUV variation \citep{dupuisetal00-1}. 

The paper is arranged as follows: In Section \ref{sec:gd394:obs} we describe the observations of GD\,394, followed by Section \ref{sec:gd394_var} where we discuss their implications for the short- and long-term variability of GD\,394. In Section \ref{sec:gd394atm} we use model atmosphere fits to the spectroscopy to measure the atmospheric parameters and metal abundances of the star. Section \ref{sec:gd394_gas} describes the search for gaseous emission from a circumstellar disc. Finally we discuss our results in Section \ref{sec:gd394_discussion} and conclude in Section \ref{sec:gd394_conclusion}.

\begin{figure}
\centering                                                            
\includegraphics[width=8.0cm]{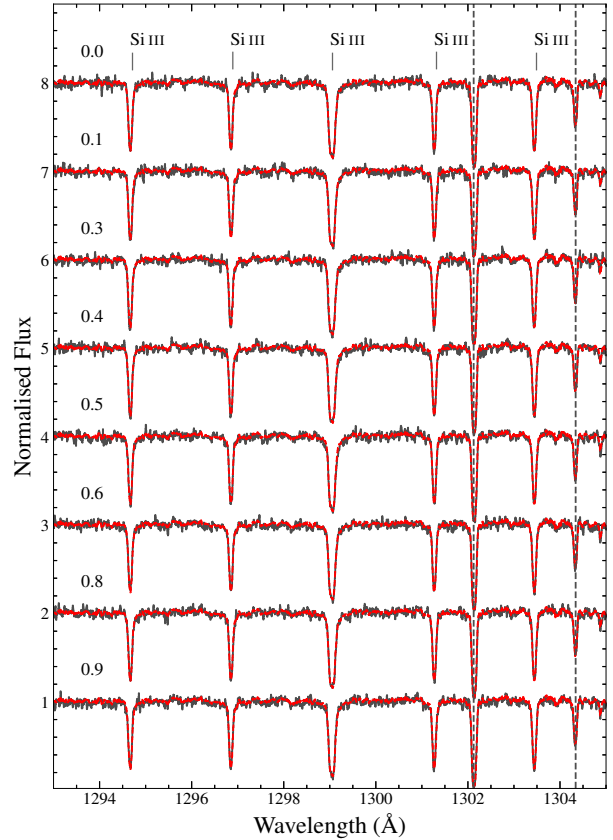}                  
\caption{Phase-resolved {\em HST}/STIS spectra of GD\,394 showing \ion{Si}{iii} lines around 1300\AA. The phase relative to the 1.15\,d period previously detected in the EUV \citep{dupuisetal00-1}, is shown on the left, with the start of the first observation defining phase zero. The co-added spectrum is over-plotted in red, illustrating the absence of variability between the observations. Interstellar \ion{O}{i} lines are indicated by the grey dashed lines. \protect\label{fig:lines_comp}}             
\end{figure} 

\begin{figure}
\centering  
\includegraphics[width=8.0cm]{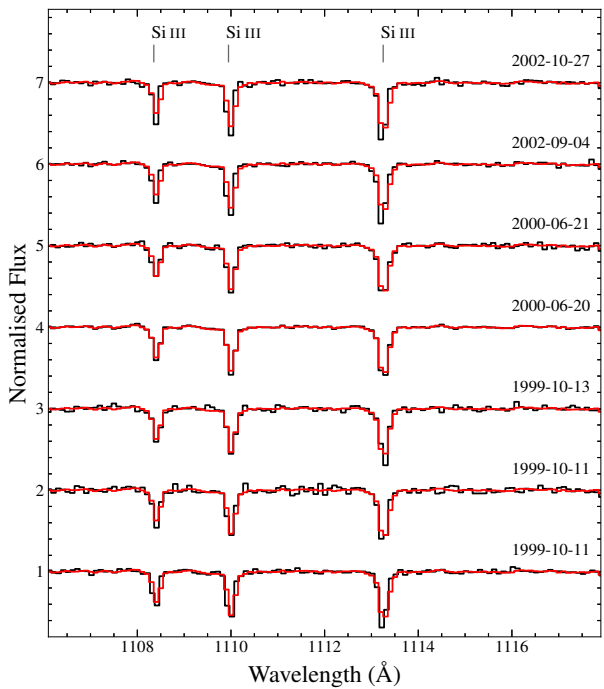}                  
\caption{Section of time-series {\em FUSE} spectroscopy of GD\,394 compared with a co-add (red), showing no significant change in absorption line strength.\protect\label{fig:fuse}}             
\end{figure}

\section{Observations}\label{sec:gd394:obs}

\subsection{Spectroscopy}\label{sec:spec}

\subsubsection{STIS Far-Ultraviolet}
To test the accretion spot hypothesis, we obtained eight far-ultraviolet (FUV) spectra of GD\,394 using the Space Telescope Imaging Spectrograph onboard the {\em Hubble Space Telescope} ({\em HST}/STIS). The observations were taken between 2015~August\,20-21, timed to fully sample the 1.15\,d period seen in the EUV. A summary of the FUV observations is given in Table \ref{tab:fuv_obs}, including their phase positions relative to the start of the first observation based on the 1.15\,d EUV period. The observations were taken with the FUV-MAMA detector in TIME-TAG mode, using the E140M grating covering the wavelength range 1144--1710\AA\ with an average resolution of $\lambda/91700$ per pixel. For each of the eight spectra we combined the echelle orders into one spectrum, co-adding at each order overlap by interpolating the order with smaller wavelength bins onto the order with wider bins\footnote{The {\sc Python} script used is available at \tt{https://github.com/davidjwilson/djw\_hst\_tools}}. The eight spectra were then co-added into one final grand average spectrum with a signal-to-noise ratio of $\approx 50-60$ and resolution of $\approx 0.02$\,\AA. Example sections of all eight spectra and the co-added spectrum are shown in Figure \ref{fig:lines_comp}.

\subsubsection{STIS Near-Ultraviolet}
GD\,394 was also observed by STIS in near-ultraviolet (NUV) as part of Program ID 13332. A spectrum was obtained on 2013\,December\,23, which used the E230H grating to cover a wavelength range of 2577--2835\,\AA\ with an average resolution of $\lambda/228,000$ per pixel. The spectrum only has S/N $\approx 6$ but clearly shows several ISM \ion{Mg}{ii} and \ion{Fe}{ii} absorption lines blue-shifted by $\approx10\mathrm{km\,s}^{-1}$.

\subsubsection{HIRES}
Optical spectroscopy was obtained on 2009~May\,23--24 and 2015~November\,15 using the High Resolution Echelle Spectrometer (HIRES) on the 10\,m W. M. Keck Telescope \citep{vogtetal94-1} under Program IDs A284Hr, A284Hb and N116Hb. On both nights $3\times300$\,s exposures were taken covering the wavelength range 3125--5997\,\AA. On the first night an additional $3\times300$\,s exposures were taken covering the wavelength range 4457--7655\,\AA\ using the GG475 filter. All of the exposures used the C5 aperture. The HIRES data were reduced using the {\sc Redux} pipeline\footnote{{\tt http://www.ucolick.org/$\sim$xavier/HIRedux/}}. Standard data reduction steps were performed including flat fielding, bias subtraction, and two-dimensional wavelength solutions produced from Th-Ar comparison exposures. The spectra were optimally extracted, with hot pixels removed using a median comparison between the individual exposures. Continuum normalization was done using low-order polynomials and ignoring any absorption lines. Spectra from adjacent orders were averaged in overlap regions, and the individual exposures at each epoch were co-added using a signal-to-noise weighted average.

The spectra contain multiple absorption lines from Si. The equivalent widths of the absorption lines are the same to within one sigma between the two epochs, so the spectra were co-added to produce a single S/N$\approx140$ spectrum.  

\subsubsection{WHT}
Further optical spectroscopy of GD\,394 was obtained using the Intermediate dispersion Spectrograph and Imaging System (ISIS) on the 4.2\,m William Herschel Telescope (WHT) on 2007~August\,6 and 2016~August\,13. The 2007 observation used the R1200 grating to cover the wavelength ranges 4520--5262\,\AA\ and 8260--9014\,\AA, with a total exposure time of 2882\,s. The 2016 observations used the R600B+R gratings covering 3056--5409\,\AA\ and 5772--9088\,\AA\ with a total exposure time of 1800\,s. The raw WHT data were reduced with standard spectroscopic techniques using {\sc starlink} software. Debiasing, flat fielding, sky-subtraction, and optimal extraction were performed using the {\sc Pamela} package. {\sc Molly} was used for wavelength and flux calibration of the 1-D spectra\footnote{{\sc pamela} and {\sc molly} were written by
  T.\ R.\ Marsh and can be obtained from {\tt http://www.warwick.ac.uk/go/trmarsh}.}. No metal absorption lines are detected in either spectrum, as the 2007 observation did not cover the appropriate wavelength ranges and the 2016 observation has insufficient spectral resolution.   

\subsection{Archival data}
In addition to the new observations described above, we utilised the following archival datasets:

GD\,394 was observed by the {\em Far Ultraviolet Spectroscopic Explorer }({\em FUSE}) spacecraft on eight occasions between 1999~October\,11 and 2002~October\,27, covering the wavelength range 925--1180\,\AA\ in the FUV. The data were retrieved from the MAST archive and recalibrated using {\sc CalFUSE v3.2.3}. The spectra from the four separate channels in the {\em FUSE} instrument were renormalised to the guiding channel and combined. Sections of each spectrum are shown in Figure \ref{fig:fuse}. As no variation between the observations was seen the eight spectra were finally coadded into one S/N $\approx 90$ spectrum. 
 
\citet{shipmanetal95-2} used the Goddard High Resolution Spectrograph(GHRS), one of the first-light instruments on {\em HST}, to observe GD\,394 on 1992\,June\,18. Spectra were taken of three sections of the FUV: Lyman\,$\alpha$, the \ion{Si}{iii} lines around 1300\,\AA, and the \ion{Si}{iv} 1392\,\AA\ and 1402\,\AA\ doublet, areas which are also covered by our STIS FUV spectrum. The data were retrieved from the MAST database.

GD\,394 was also observed multiple times by the {\em International Ultraviolet Explorer} ({\em IUE}) in both high and low resolution mode. A full description and analysis of the high-resolution {\em IUE} spectra is presented in \cite{holbergetal98-1}.

Finally, we retrieved an optical spectrum obtained by \citet{gianninasetal11-1} from the Montreal White Dwarf Database \citep[MWDD,][]{dufouretal17-1} covering the wavelength range 3780--5280\,\AA.\\

A full list of all detected metal absorption lines across the full wavelength range covered by our spectra is given in Appendix \ref{sec:line_lists}.

\begin{figure}
    \centering
    \includegraphics[width=8.0cm]{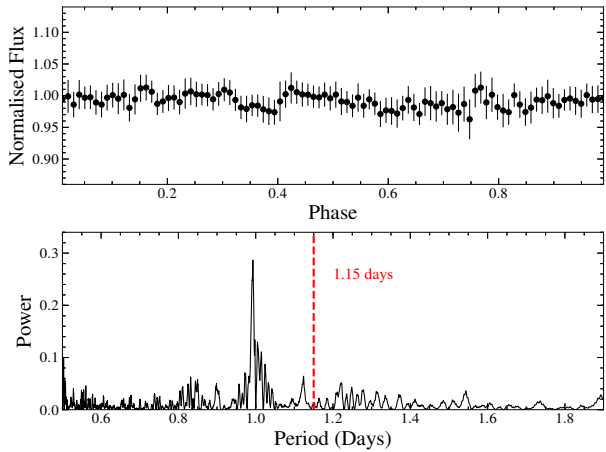}
    \caption{Observations of GD\,394 made by SuperWASP. Top: SuperWASP data binned to 1000\,s and folded on a 1.15\,d period. Bottom: Lomb-Scargle periodogram, showing no significant peaks apart from a 24 hour alias.      \protect\label{fig:swasp}}. 
\end{figure}

\subsection{Photometric observations}\label{sec:lc}

\subsubsection{SuperWASP}
GD\,394 was observed by the Wide-Angle Search for Planets (SuperWASP) on multiple occasions between 2006\,November and 2008\,August \citep{faedietal11-1}. 3506 30\,s exposures were taken with a median S/N$=7.5$. The data were retrieved from the WASP archive and calibrated using the standard routines described in \citet{pollaccoetal06-1}.

\subsubsection{W1m}
Additional photometry was obtained using the Warwick 1m telescope (W1m) telescope at the Roque de Los Muchachos Observatory on La Palma on 2016\,May\,14 and 2016\,May\,16. The telescope has a dual beam camera system with fixed visual and $Z$ band filters and was operated in engineering mode. A cadence of 13\,s was achieved using 10\,s exposures and a 3\,s readout time. Total time on target was 2990\,s and 7970\,s on the two nights respectively, covering 11\,per\,cent of the 1.15\,d EUV period.

\subsubsection{Ultraviolet photometry}
Finally, both the STIS and {\em FUSE} spectroscopic observations were obtained using photon-counting detectors , allowing high-cadence lightcurves to be obtained \citep[see for e.g.][]{wilsonetal15-1, sandhausetal16-1}.  

The STIS lightcurve was extracted from the TIME-TAG event files with events from the edge of the detector and around the Ly\,$\alpha$ line removed. As GD\,394 is relativity bright and the echelle spectral trace covers most of the detector, no background subtraction was necessary. Each lightcurve contains multiple irregular flux dropouts of approximately one second. Comparison with the jitter (\textsc{.jit}) files, which record the precise spacecraft pointing during the exposure, showed that each dropout was accompanied by a sharp spike in pointing declination. All times associated with spikes in declination were therefore masked out, regardless of whether flux dropouts were seen.    

Each lightcurve is dominated by a semi-linear increase in flux over the exposure of $\simeq5$\,per\,cent. As this trend is seen in every observation, regardless of phase position within the 1.15\,d EUV period of GD\,394, it is likely instrumental in origin. ``Breathing'' of the telescope induced by the changing thermal environment experienced by {\em HST} during its orbit has been shown to alter the focus position, with STIS requiring roughly one orbit after a change in pointing to thermally relax \citep{singetal13-1}. As our visits are only one orbit long we cannot use the usual method of discarding data from the beginning of the visit. Dividing by a linear fit removes most of the trend, but still leaves variations in flux that preclude an accurate measurement of variability below the $\approx 1$\,per\,cent level. However, we can rule out any changes above that level over the 1.15\,d period.
 
Light-curves from the {\em FUSE} observations were extracted in a similar fashion, providing a useful link between the STIS FUV and {\em EUVE} observations. Unfortunately the observations are affected by aperture drift caused by thermal distortion, so absolute photometry is impossible and we cannot compare between observations. Nevertheless, each individual light-curve is constant, with no evidence for flux variations similar to those detected in the EUV.

\subsection{{\em Gaia} parallax}
Astrometric measurements of GD\,394 by the {\em Gaia} spacecraft were included in {\em Gaia} Data Release 2 \citep{gaia18-1}. We queried the {\em Gaia} archive for GD\,394, finding a unique match with source ID 2166111956258599680 and parallax $\varpi = 19.85\pm0.064$\,mas. The precision of the parallax measurement is sufficiently high that it can be directly converted into a distance \citep{bailer-jonesetal18-1}, placing GD\,394 at $d=50.37\pm0.16$\,pc.

\begin{table*} 
\centering 
\caption{Equivalent widths of the lines shown in Figure \ref{fig:lines_comp} across all eight {\em HST} visits.} 
\begin{tabular}{llccccccccc}\\ 
\hline 
& & \multicolumn{8}{c}{Equivalent Width (m\AA)} \\ 
Line & Rest $\lambda$ (\AA)& Visit 1 & Visit 2 & Visit 3 & Visit 4 & Visit 5 & Visit 6 & Visit 7 & Visit 8\\ \hline 
\ion{Si}{iii} & 1294.54 & $77.0\pm7.8$ & $86.0\pm7.7$ & $79.0\pm7.6$ & $83.0\pm7.7$ & $80.0\pm7.7$ & $79.0\pm7.8$ & $82.0\pm7.1$ & $78.0\pm7.8$\\  
\ion{Si}{iii} & 1296.73 & $77.0\pm7.6$ & $79.0\pm8.0$ & $76.0\pm7.5$ & $79.0\pm7.7$ & $78.0\pm7.3$ & $80.0\pm7.9$ & $77.0\pm7.2$ & $77.0\pm8.1$\\  
\ion{Si}{iii} & 1298.89 & $150\pm11$ & $150\pm12$ & $150\pm12$ & $150\pm12$ & $160\pm13$ & $150\pm12$ & $150\pm12$ & $150\pm12$\\  
\ion{Si}{iii} & 1301.15 & $76.0\pm7.1$ & $71.0\pm7.1$ & $74.0\pm7.1$ & $81.0\pm7.6$ & $75.0\pm7.4$ & $77.0\pm7.1$ & $79.0\pm7.3$ & $72.0\pm7.2$\\  
\ion{Si}{iii} & 1303.32 & $77.0\pm6.9$ & $80.0\pm7.3$ & $78.0\pm7.5$ & $75.0\pm7.4$ & $81.0\pm7.8$ & $75.0\pm7.4$ & $80.0\pm7.2$ & $72.0\pm7.1$\\  
\ion{Si}{iii} & 1312.59 & $47.0\pm6.4$ & $43.0\pm6.1$ & $45.0\pm6.9$ & $43.0\pm6.5$ & $45.0\pm6.0$ & $46.0\pm6.7$ & $47.0\pm6.1$ & $48.0\pm6.4$\\  
\hline 
\end{tabular} 
\label{tab:ews} 
\end{table*}

\section{Non-detection of variability} \label{sec:gd394_var}

\subsection{Short-term}
Figure \ref{fig:lines_comp} shows sections of all eight STIS FUV spectra, compared with the co-add of all eight spectra. The spectra show neither significant changes in the strength, shape or velocity of the absorption lines, nor any change in the total flux or flux distribution. Equivalent widths for all identifiable lines were measured via the formalism given by \citet{vollmann+eversberg06-1}, all of which were constant to within one sigma. The equivalent widths of the lines shown in Figure \ref{fig:lines_comp} are given in Table \ref{tab:ews}. We thus detect no evidence for an accretion spot in the FUV. 

The {\em FUSE} spectra also show no signs of variability (Figure \ref{fig:fuse}). By coincidence, the {\em FUSE} observations sample the 1.15\,d EUV period fairly evenly, although their spacing in time is too large to precisely map their phase position. Therefore, they also provide no evidence for an accretion spot, although they are a weaker constraint compared to the STIS FUV observations.  

The second consequence of the accretion spot hypothesis put forward by \citet{dupuisetal00-1} is that GD\,394 should undergo a periodic variation in optical flux, in anti-phase with the EUV variation, due to flux redistribution. Figure\,\ref{fig:swasp} shows our SuperWASP lightcurve folded onto the 1.15\,d EUV period, along with a Lomb-Scargle periodogram. No significant variation is seen in the lightcurve, and the periodogram does not return a 1.15\,d period. Neither do the W1m or ultraviolet lightcurves show any variation consistent with a 1.15\,d period. This strengthens the conclusion drawn from the lack of spectral variation that the spot hypothesis is incorrect. Additionally, no evidence for transiting debris like that at WD\,1145+017 \citep{vanderburgetal15-1} is detected in any of the available photometry. 

\subsection{Long-term}
A possible explanation for the lack of short-term variability is that the accretion spot dispersed in the time between the {\em EUVE} and more recent observations. We can test this via a comparison between the two available epochs of {\em HST} FUV spectroscopy, obtained with GHRS in 1992 and with STIS in 2015. As the GHRS spectrum was obtained within two years of the {\em EUVE} observations (which spanned 1993--1996), it is reasonable to assume that the EUV variation was already present. If a spot existed in 1992 but not in 2015, then the strength of the absorption lines should likely be different. We find that equivalent widths of the absorption lines detected by both instruments are identical to within 1\,sigma (Table \ref{tab:ghrsews}). The variation must therefore either still have been present when our STIS observations were obtained or have stopped in such a way as to leave the FUV spectrum unchanged, unless by an unlikely coincidence the GHRS spectra were obtained at a phase where the varying absorption lines were at the same strength as the later lines with constant strength. The lack of change over a 23 year period joins the long term observations of several other accreting white dwarfs \citep{wilsonetal14-1, manseretal16-1} where no changes in absorption line strength have been detected. Given that the metal diffusion timescales are much shorter than the time between observations, the accretion rates onto metal-polluted white dwarfs are remarkably stable.  

We note that, although the strength of the absorption lines is constant, the continuum fluxes of the GHRS and STIS FUV spectra differ. Specifically, the GHRS data around Ly\,$\alpha$, 1300\,\AA, and 1400\,\AA\ have median fluxes $\approx10$\,per\,cent higher, similar, and $\approx10$\,per\,cent lower than the corresponding STIS FUV observations, respectively. Such a change in the continuum flux of GD\,394 is physically implausible, and is probably related to the issues with the STIS flux calibration discussed in Section \ref{sec:stis_mod}.

\begin{table} 
\centering 
\caption{Equivalent widths of absorption lines detected in the GHRS spectra compared with the co-added STIS spectrum.} 
\begin{tabular}{llcc}\\ 
\hline 
& & \multicolumn{2}{c}{Equivalent Width (m\AA)} \\ 
Line & Rest $\lambda$ (\AA)& GHRS\,(1992) & STIS\,(2015)  \\ \hline
\ion{Si}{ii} & 1206.5 & $ 270.0 \pm 170.0 $ & $ 260.0 \pm 73.0 $ \\
\ion{Si}{iii} & 1207.517 & $ 16.0 \pm 45.0 $ & $ 32.0 \pm 19.0 $ \\
\ion{Si}{iii} & 1294.545 & $ 68.0 \pm 27.0 $ & $ 81.0 \pm 5.7 $ \\
\ion{Si}{iii} & 1296.726 & $ 77.0 \pm 47.0 $ & $ 77.0 \pm 5.0 $ \\
\ion{Si}{iii} & 1298.892 & $ 140.0 \pm 48.0 $ & $ 150.0 \pm 7.9 $ \\
\ion{Si}{iii} & 1301.149 & $ 68.0 \pm 35.0 $ & $ 77.0 \pm 5.3 $ \\
\ion{Si}{iii} & 1303.323 & $ 88.0 \pm 44.0 $ & $ 74.0 \pm 5.6 $ \\
\ion{Si}{iii} & 1312.591 & $ 43.0 \pm 42.0 $ & $ 45.0 \pm 6.3 $ \\
\ion{Si}{iv} & 1393.755 & $ 490.0 \pm 130.0 $ & $ 490.0 \pm 41.0 $ \\
\ion{Si}{iv} & 1402.77 & $ 340.0 \pm 100.0 $ & $ 320.0 \pm 20.0 $ \\

\hline 
\end{tabular} 
\label{tab:ghrsews} 
\end{table} 

\begin{table*} 
\centering 
\caption{Radial velocity measurements of GD\,394 over all available epochs, adjusted to the zero-point of the STIS FUV spectra using ISM lines. Velocities given in $\mathrm{km\,s}^{-1}$.} 
\begin{tabular}{lccccc}\\ 
\hline 
Observation & Epoch & v$_{\mathrm{Photosphere}}$ & v$_{\mathrm{ISM}}$ & $\Delta$\,v & v$_{\mathrm{Adjusted}}$ \\
\hline
SWP high resolution &	$1988.9\pm 6.6$ & $28.9\pm0.8$	& $-6.18\pm1.46$ &	$-2.88$ &	$26.02\pm1.5$ \\
GHRS &	$1992.464$ & $35.37\pm2.81$	& $-0.4\pm3.18$ &	$-8.66$	& $26.71\pm2.3$\\
Lick &	$1996.68$ &	$27.6\pm1.3$ & - &	- &	$27.6\pm1.3$\\
HIRES &	$2009.39$ &	$28.49\pm1.3$ &	$-10.59\pm1.2$ &	$1.53$ & $30.02\pm1.8$\\	
HIRES &	$2015.87$ & $27.53\pm1.2$ &	$-11.06\pm0.88$	& $2.00$ &	$29.53\pm1.5$\\	
STIS &	$2015.64$ & $27.66\pm0.36$ & $-9.06\pm0.2$	& $0.00$ &	$27.66\pm0.36$\\
\hline 
\end{tabular} 
\label{tab:rvs} 
\end{table*}

We can also place upper limits on long-term changes in the velocities of the stellar features seen in GD\,394. Between 1982 and 2015 various high-dispersion observations have been obtained, including: (1) {\em International Ultraviolet Explorer} ({\em IUE}) high dispersion spectra obtained between 1982 and 1994 \citep{holbergetal98-1}; (2) {\em HST}/GHRS spectra obtained in 1992 \citep{shipmanetal95-2}; (3) Lick, Mt. Hamilton Observatory spectra in 1996 \citep{dupuisetal00-1}; (4) Keck/HIRES observations in 2009 and 2015 (this paper); and (5) {\em HST}/STIS data in 2015 (this paper). These various observations, obtained in the vacuum ultraviolet and the optical, potentially have different wavelength scales and velocity zero points. However, by using interstellar lines as an invariant fiducial, we can place these observations on the velocity scale of the STIS data.

$\bullet$ {\em IUE} high resolution SWP spectra: Four observations were obtained in 1982 May 05, 1984 April 15, 1994 January 03 and 1994 January 04, respectively. Because they were all obtained through the SWP large aperture they included wavelength offsets due to location of the stellar image within the aperture. The process of co-adding these spectra involved small wavelength displacements applied to each spectrum \citep[see][]{holbergetal98-1}, which were used to co-align the observed stellar and interstellar lines in each spectrum to an arbitrary zero-point prior to co-addition. A detailed examination of this process shows that there were no relative displacements between ISM and stellar lines for the individual spectra above the 10\,m\AA\ level, or approximately $\approx 2\,\mathrm{km\,s}^{-1}$. Thus, we detect no evidence of velocity variations both prior to and during the time span when the {\em EUVE} 1.15\,d variations were observed.

$\bullet$ {\em HST}/GHRS spectra: The three ultraviolet spectral bands of GD\,394 observed on 1992 June 18 by \citep{shipmanetal95-2} included both ISM and photosphere lines. \citet{dupuisetal00-1} re-measured the GHRS spectra, noting that wavelength calibrations were not obtained for one of the grating settings. We also re-measured the GHRS spectra using the same software as used for the IUE spectra. We find photospheric and ISM velocities of $35.37\pm2.81\mathrm{km\,s}^{-1}$ and $-0.40\pm3.18\mathrm{km\,s}^{-1}$ respectively.

$\bullet$ Lick Mt. Hamilton Echelle: On 1996 September 6--7 \citet{dupuisetal00-1} observed  photospheric \ion{Si}{iii} features in GD\,394 and report a velocity of $27.6\pm1.3\mathrm{km\,s}^{-1}$. No interstellar lines were reported and hence we take this velocity at face value.

$\bullet$ Keck/HIRES: Two Keck HIRES spectra (see Section~\ref{sec:gd394:obs}) were obtained in 2009 and 2015 and contain photospheric \ion{Si}{iii} and interstellar \ion{Ca}{iii} absorption lines. Seven individual \ion{Si}{iii} lines were measured in the 2015 data but only five were measurable in the 2009 data. The interstellar Ca\,K line is measurable in both spectra, whilst the Ca\,H line is barely seen in the 2015 data and is not detected in the 2009 data.

$\bullet$ {\em HST}/STIS: The photospheric and interstellar velocities measured from the STIS FUV spectra obtained in August 2015 represent averages from Tables \ref{tab:gd394_ps_lines} and \ref{tab:gd394_cs_lines}.

In Table \ref{tab:rvs} we list the instruments,epochs and measured observed photospheric and ISM velocities, where available. Also listed are the velocity adjustments ($\Delta v$) necessary to align the ISM velocities of the different instruments with those of the STIS data. The final column of Table \ref{tab:rvs} gives these adjusted photospheric velocities. There is no evidence of any significant radial velocity variation of GD\,394 between 1982 and 2015. 


\begin{table}
\centering
\caption{Atmospheric parameter determinations for GD\,394 from the literature. References: 1.\,\citet{holbergetal86-1}; 2.\,\citet{finleyetal90-1}; 3.\,\citet{kidderetal91-1}; 4.\,\citet{vennes92-1}; 5.\,\citet{bergeronetal92-1}; 6.\,\citet{barstowetal96-1}; 7.\,\citet{marshetal97-1}; 8.\,\citet{vennesetal97-1}; 9.\,\citet{finleyetal97-1}; 10.\,\citet{dupuisetal00-1}; 11.\,\citet{lajoie+bergeron07-1}; 12.\,\citet{gianninasetal11-1}; 13.\, This work.}
\begin{tabular}{r@{\,$\pm$\,}lr@{\,$\pm$\,}lcc}\hline
\multicolumn{2}{c}{$T_{\mathrm{eff}}$\,(K)} & 
\multicolumn{2}{c}{$\log g$\,(cm\,s$^{-2}$)} & Data & Ref. \\
\hline
36125 & 940             & 8.13 & 0.25	  & Ly\,$\alpha$          & 1  \\
36910 & $1630\atop1410$ & \multicolumn{2}{c}{8.00 (fixed)}               & FUV	            & 2  \\
39800 & 1100            & 8.05 & 0.31	  & Ly\,$\alpha$/Balmer  & 3  \\
37000 & 1500            & 8.25 & 0.25     & EUV/Ly\,$\alpha$/FUV & 4  \\
39450 & 200	            & 7.83 & 0.04	  & Balmer	        & 5  \\
40300 & $400\atop500$   & 7.99 & $+0.08 \atop -0.05$  & Ly\,$\alpha$/Balmer  & 6  \\
38866 & 730	            & 7.84 & 0.10     & Balmer	        & 7  \\
39800 & 300	            & 8.00 & 0.04	  & Balmer	        & 8  \\
39639 & 40              & 7.938&0.027	  & Balmer          & 9 \\
35044 & 25              & 7.86 & 0.02	  & IUE/HUT/GHRS    & 10 \\
39205 & 470             & 7.81 & 0.038    & Balmer          & 11 \\
32788 & 3800            & 7.81 & 0.038    & UV (IUE)        & 11 \\  
34750 & 2575            & 7.81 & 0.038    & UV/V            & 11 \\
39660 & 636             & 7.88 & 0.05     & Balmer & 12 \\[1ex]
35700 & 1500            & 8.05 & 0.20  & STIS+{\em Gaia}+phot.   & 13 \\
41000 & 1000            & 7.93 & 0.10  & Balmer & 13\\  
\hline
\end{tabular}
\label{tab:past_atm}
\end{table}

\begin{table}
\centering
\caption{Photometric magnitudes from Pan-STARRS and 2MASS.}
\begin{tabular}{l l | l l}
\hline
&Pan-STARRS & &2MASS\\
\hline
$r$    & 13.345$\pm$0.0010  & $J$     & 13.755$\pm$0.0330 \\
$i$    & 13.698$\pm$0.0010 & $H$     & 13.791$\pm$0.0450 \\
$z$    & 13.963$\pm$0.0005   & $K_\mathrm{s}$ & 13.982$\pm$0.0530 \\
$y$    & 14.157$\pm$0.0050 & &\\

\hline
\end{tabular}
\label{tab:photometry}
\end{table}

\begin{table} 
\centering 
\caption{Characteristics of GD\,394 for the different atmospheric parameters measured from ultraviolet and optical spectra, computed using the Evolutionary Tables tool on the MWDD.}
\begin{tabular}{lll}\\ 
\hline 
& Ultraviolet & Visible \\
\hline 
$T_{\mathrm{eff}}$\,(K) & $35700\pm1500$   &  $41000 \pm2000$ \\
$\log g$\,(cm\,s$^{-2}$)         & $8.05\pm0.20$    & $7.93 \pm0.10$ \\
Mass (\Msun)            & $0.69\pm0.10$   & $0.639\pm0.048$\\
Radius (\Rsun)          & $0.013\pm0.002$ & $0.0143\pm0.0011$\\
Cooling age (Myr)      & $6.0\pm0.1$     & $4.0\pm0.5$\\
\hline
\end{tabular} 
\label{tab:characteristics} 
\end{table}

\section{Atmospheric parameters and metal abundances}\label{sec:gd394atm}

\begin{figure}
    \centering
    \includegraphics[width=1\columnwidth]{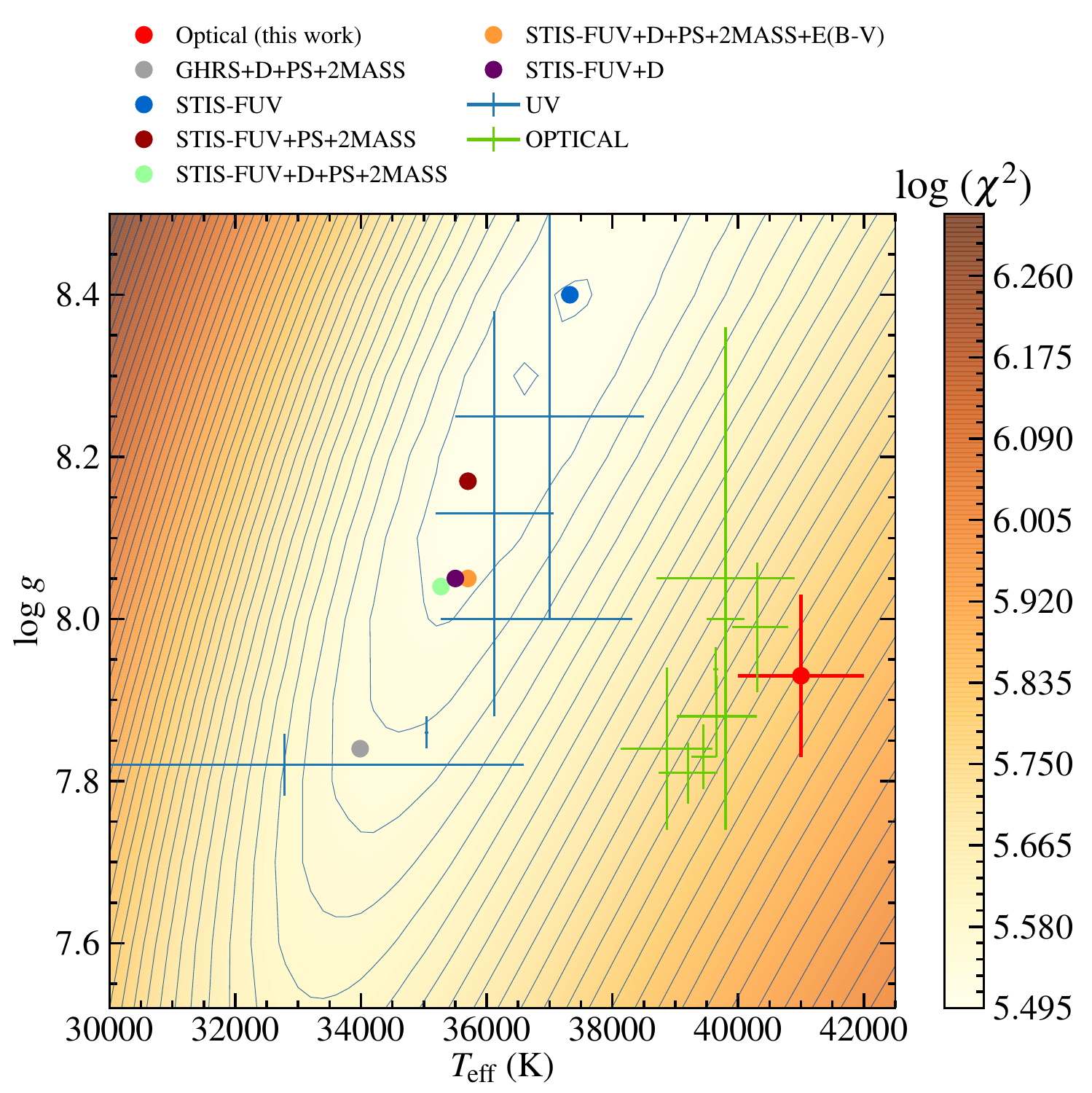}
    \caption{Atmospheric parameter fits to GD\,394. The underlying contour plot shows the $\log\chi^2$ space of the fit to the Ly\,$\alpha$ line without any additional constraints. Results for fits including priors (\textit{Gaia} distance $D$, Pan-STARRS (PS) and 2MASS photometry) and interstellar extinction, $E(B-V)$, are over-plotted as coloured dots, with our final adopted values shown in orange. The blue and green markers show the literature determinations from Table\,\ref{tab:past_atm} for fits to ultraviolet and optical spectroscopy, respectively.\label{fig:contour}}
\end{figure}

\begin{figure}
    \centering
    \includegraphics[width=1\columnwidth]{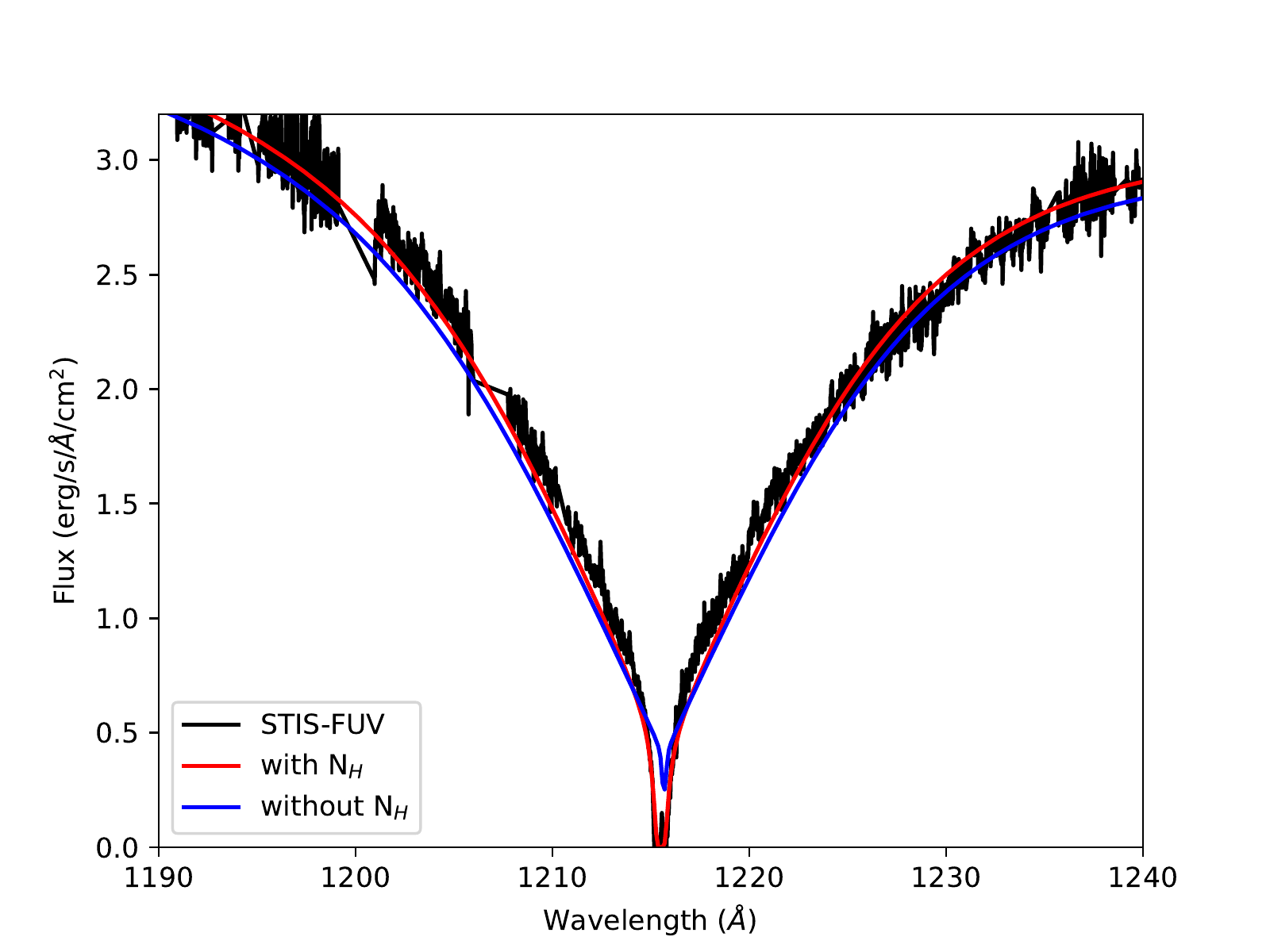}
    \caption{STIS FUV spectrum of the Ly\,$\alpha$ core, with the best-fitting model without (blue, $T_\mathrm{eff}=35273, \log g=8.036$) and with (red, $T_\mathrm{eff}=35700, \log g=8.054$) interstellar extinction over-plotted, demonstrating the improvement to the fit when absorption by $\mathrm{N_H}=2.6\times10^{18}\mathrm{cm^{-2}}$ neutral hydrogen is included. The metal lines were masked out for this fit. \label{fig:lyman}} 
\end{figure}

\begin{figure*}
    \centering
    \includegraphics[width=\textwidth]{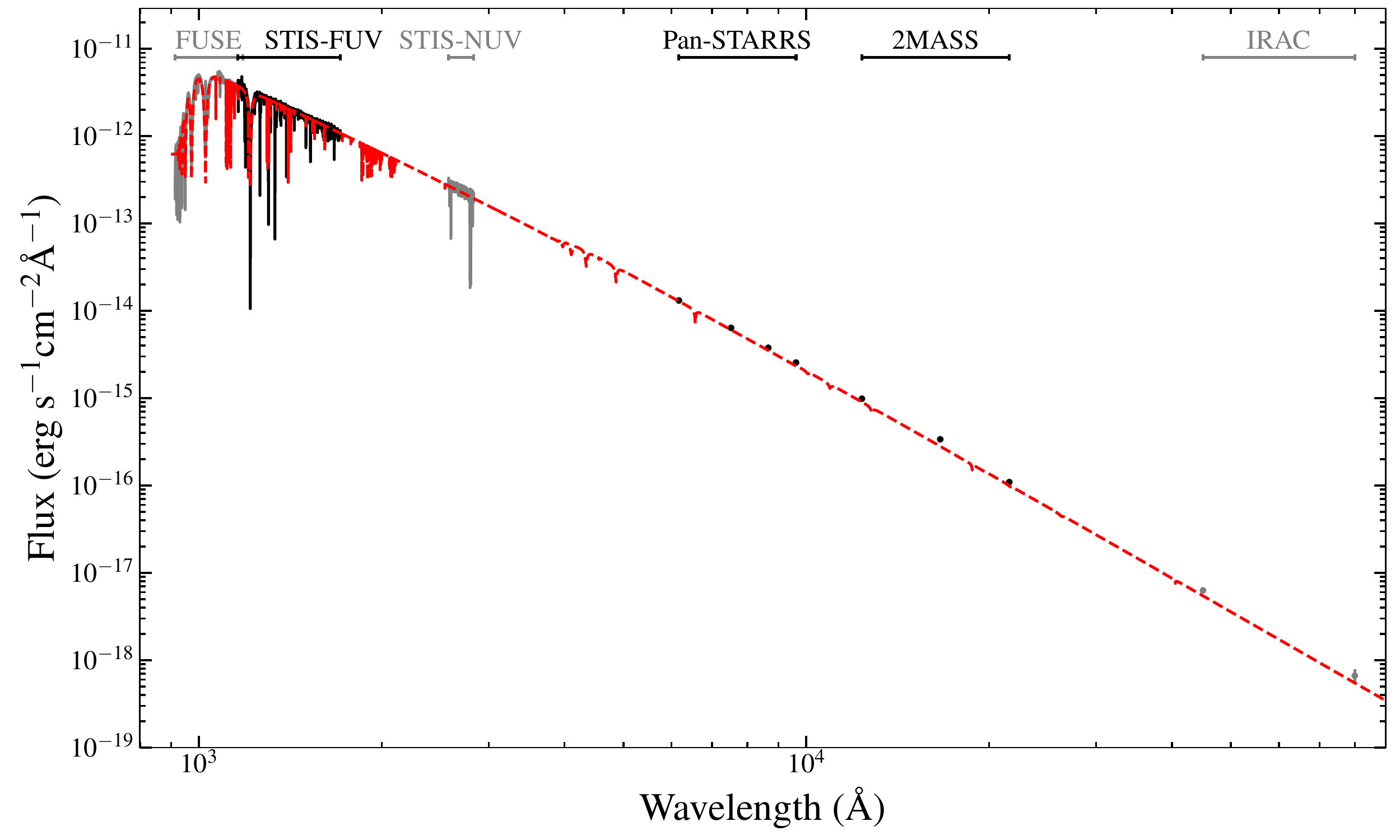}
    \caption{Spectral energy distribution of GD\,394 from the far ultraviolet to mid-infrared, over-plotted with the model fit based on found using the STIS FUV spectroscopy, the {\em Gaia} parallax and the Pan-STARRS and 2MASS photometry. The model was scaled to the STIS FUV spectrum for the plot. Data in black was used in the fitting process, whilst data in grey was not fitted but serves as further confirmation of the model. The {\em Spitzer}/IRAC data are taken from \citet{mullallyetal07-1}. \protect\label{fig:sed}} 
\end{figure*}

\begin{figure}
    \centering
    \includegraphics[width=8.cm]{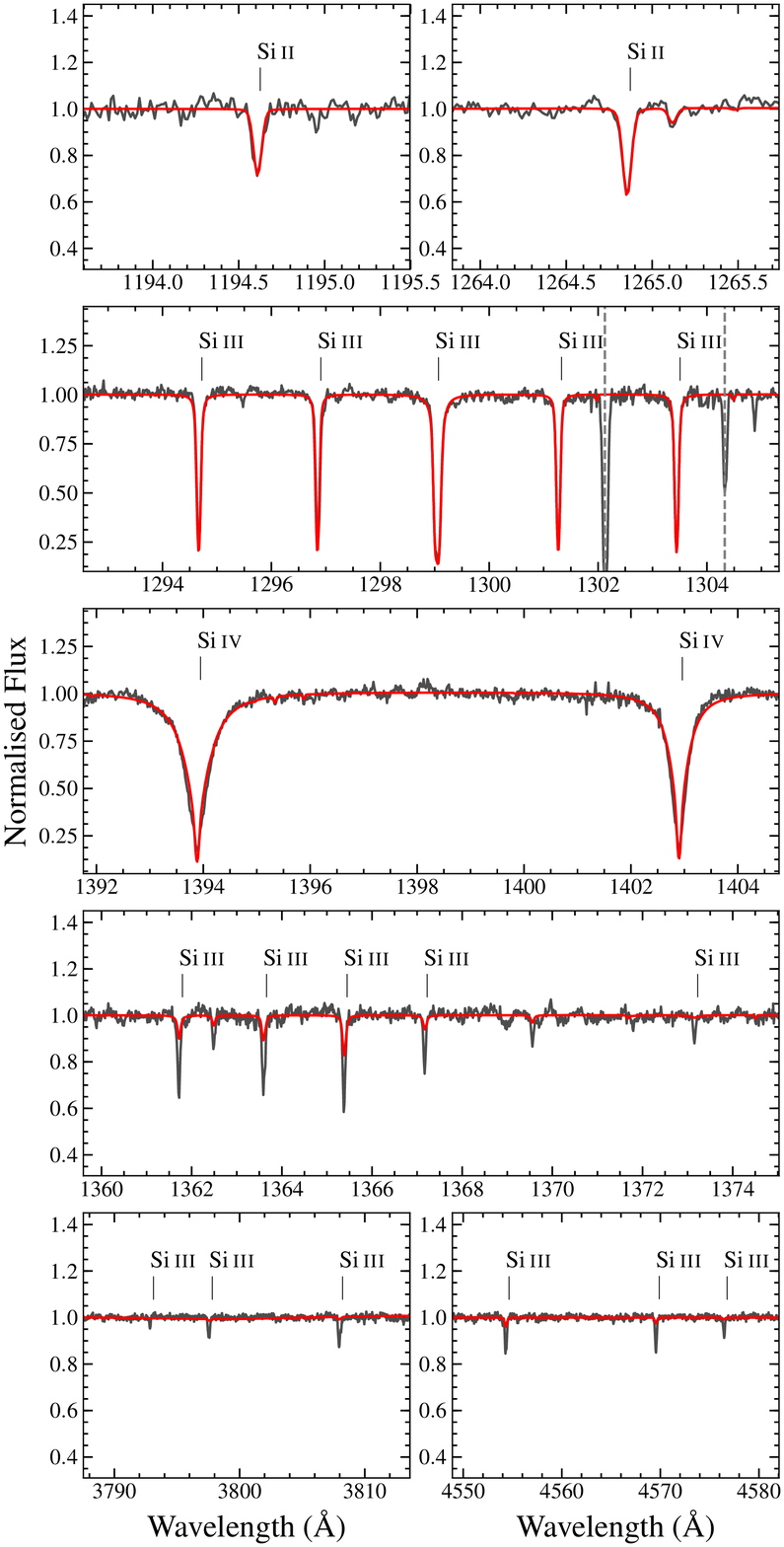}
    \caption{Model atmosphere (red) to various ionisation levels of Si. \ion{Si}{ii}, \ion{Si}{iv} and low-excitation states of \ion{Si}{iii} (top three panels) are all consistent with and abundance of Si/H$\approx -5.9$. \ion{Si}{iii} lines with excitation energies $\gtrsim 7$\,eV instead require Si/H$\approx -5.1$ in both the STIS and HIRES spectra (bottom two panels).   \protect\label{fig:gd394_si_levels}}
\end{figure}

\begin{figure*}
    \centering
    \includegraphics[width=16.0cm]{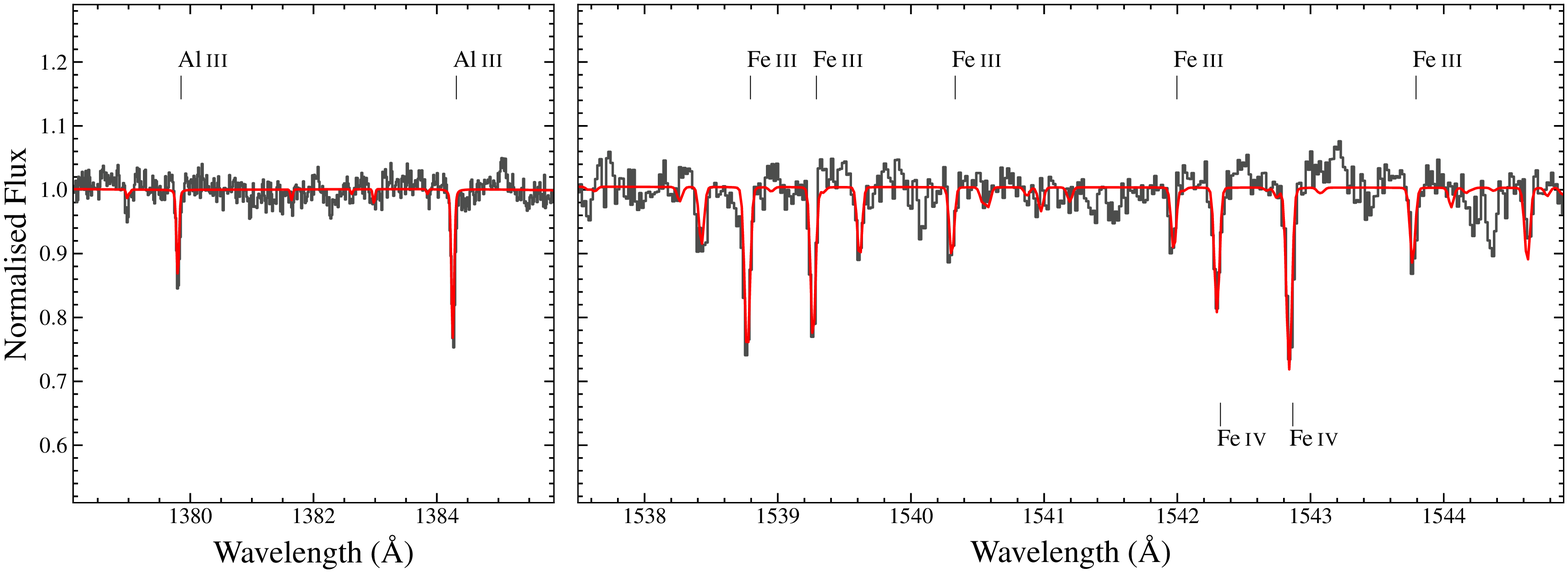}
    \caption{Model fit (red) to example Al and Fe absorption lines in the co\bs{-}added STIS FUV spectrum. \protect\label{fig:gd394_alfe}} 
\end{figure*}

Previously published estimates of the atmospheric parameters of GD\,394 are collected in Table \ref{tab:past_atm}. Apparent in these results is a consistent discrepancy between temperatures derived from optical and ultraviolet spectroscopy, with the latter being typically cooler by $\approx$4000\,K. Similar discrepancies between fits to optical and ultraviolet spectra are seen at several hot white dwarfs, although GD\,394 is the most pronounced case \citep{lajoie+bergeron07-1}. Here we fit our extensive spectroscopic data with the latest version of the LTE white dwarf model atmosphere code described in \citet{koester10-1},  but with updated input physics, including among other data the improved hydrogen Stark broadening calculations of \citet{tremblay+bergeron09-1} and Tremblay (2015, priv.comm.)

\subsection{Optical spectroscopy}
Neither the WHT spectrum nor the spectrum from \citet{gianninasetal11-1} retrieved from the MWDD are ideal for model atmosphere fitting due to poor flux calibration and low resolution, respectively. Nevertheless, fitting to the Balmer lines returns $T_{\mathrm{eff}}=41387\pm32$\,K, $\log g=7.927\pm0.003$ and $T_{\mathrm{eff}}=39082\pm100$\,K, $\log g=8.022\pm0.016$, respectively (statistical uncertainties only), consistent with the results from the literature detailed in Table \ref{tab:past_atm}. We therefore adopt Balmer line parameters by averaging the results and using the difference as an estimate of the systematic errors, $T_{\mathrm{eff}}=41000\pm2000$\,K and $\log g=7.93\pm0.10$.

\subsection{STIS}\label{sec:stis_mod}
The STIS FUV spectra were heavily affected by ripples caused by incorrect calibration of the echelle blaze function (see STIS ISR 2018-01\footnote{\tt{http://www.stsci.edu/hst/stis/documents/isrs/2018\_01.pdf}} for a detailed discussion), and we found that the choice of the echelle blaze function (\textsc{PHOTTAB})used for the calibration of the E140M data significantly changes the best-fitting parameters measured from the Ly\,$\alpha$ line. In the following analysis we used the latest (at time of writing) calibration files, detailed in the 2018 July STScI Analysis Newsletter\footnote{\tt{http://www.stsci.edu/hst/stis/documents/newsletters/\\stis\_newsletters/2018\_07/stan1807.html\#article2}}. The artefacts remaining in the calibration do not visually affect the Ly\,$\alpha$ line, but we nevertheless caution that future improvements to the STIS calibration may require the atmospheric parameters to be reappraised.

We fitted the STIS FUV spectrum using a grid of pure hydrogen models covering the temperature range $30000$\,K $\leq T_{\mathrm{eff}} \leq 45000$\,K in steps of 200\,K, and surface gravities $7.00\leq \log g \leq 8.50$ in steps of 0.1\,dex. The metal absorption lines were masked out during the fitting process. As the Ly\,$\alpha$ profile is sensitive to the degree of ionisation of hydrogen and Stark broadening, both of which increase with the effective temperature, there is a strong correlation between these two parameters. The colour intensity map in Figure \ref{fig:contour} shows an extended valley of low $\chi^{2}$ in the $T_{\mathrm{eff}} - \log g$ space, and the STIS FUV data is fitted nearly equally well by all of these solutions. To lift this degeneracy, we introduce prior constraints to the fit using the MCMC ensemble sampler {\sc emcee} \citep{foreman-mackeyetal13-1}. The normalisation parameter that scales the model to the observations (which depends on the distance to and radius of the star) is constrained using the distance inferred from the {\em Gaia} parallax of 50.37$\pm$0.16\,parsecs combined with a mass-radius relation for DA white dwarfs\footnote{
  {\tt http://www.astro.umontreal.ca/$\sim$bergeron/CoolingModels}, based on
  \cite{holberg+bergeron06-1, kowalski+saumon06-1,
    tremblayetal11-2,bergeronetal11-1}.}. The continuum slope of the synthetic model is constrained using photometric data at longer wavelengths (specifically Pan-STARRS $r$, $i$, $z$, and $y$ \citep{chambersetal16-1} and 2MASS $J$, $H$, and $K_\mathrm{s}$   \citep{skrutskieetal06-1} magnitudes, Table \ref{tab:photometry}, with synthetic magnitudes computed by convolving the white dwarf model with the transmission function of the corresponding filters. Figure \ref{fig:contour} illustrates the change in best-fitting parameters with increasing number of constraints.
    
At the distance of GD\,394 the effects of interstellar extinction are expected to be small, but for completeness, we include $E(B-V)$ as a free parameter in the fits. The core of Ly\,$\alpha$ shows clear absorption of interstellar neutral hydrogen ($\mathrm{N_H}$)  with a radial velocity of $-13.74\pm5.28\,\mathrm{km\,s^{-1}}$, in agreement with the detected ISM metal absorption lines (Table \ref{tab:gd394_ism_lines}), and including $\mathrm{N_H}$ with a column density of $(2.618 \pm 0.044) \times 10^{18}$ atoms cm$^{-2}$ significantly improves the fit to the core of Ly\,$\alpha$ (Figure\,\ref{fig:lyman}). The column density of $\mathrm{N_H}$ is linearly correlated to the reddening \citep{diplas+savage94-1}, corresponding to $E(B-V)=(5.31\pm0.09)\times10^{-4}$, producing a final result for the atmospheric parameters of GD\,394 of $T_{\mathrm{eff}}=35700\pm12$\,K, $\log g=8.054\pm0.001$ (statistical uncertainties only). In principle, the reddening could be further constrained by fitting to available 2MASS and {\em Spitzer} photometry \citep{mullallyetal07-1}, but the value of $E(B-V)$ estimated from the $\mathrm{N_H}$ absorption in the core of Ly\,$\alpha$ is so
small that the uncertainties on the photometry (both statistical and
systematic) do not warrant this exercise. The full spectral energy distribution of GD\,394 is shown in Figure\,\ref{fig:sed}, demonstrating good agreement with the model at all wavelengths. 

\subsection{GHRS}
As noted above, the flux calibration of the GHRS and STIS FUV data are different. We refit the GHRS spectra using the same process as for the STIS FUV, again incorporating the {\em Gaia} parallax and Pan-STARRS and 2MASS photometry. We find atmospheric parameters of $T_{\mathrm{eff}}=33986\pm17$\,K and $\log g=7.841\pm0.005$. As the GHRS data is inferior to the STIS FUV data in both wavelength coverage and S/N we do not present this as an alternative result for the atmospheric parameters, but it provides a guide to estimate the systematic uncertainties of the model fit.

\subsection{FUSE}\label{sec:fuse_mod}
The absolute flux of the {\em FUSE} spectrum agrees well with the STIS spectrum in the region of overlap. Extending the model fit to the STIS spectrum to shorter wavelengths, we find that it also agrees well with the {\em FUSE} data without any changes in $T_{\mathrm{eff}}$ or $\log g$, except for the highest Lyman lines where interstellar absorption becomes dominant. The overall fit to the Si and Fe lines (the only ones we detect) is not perfect, but satisfactory. We have not attempted to re-fit the abundances, as the STIS spectrum has superior resolution and atomic data that is likely more accurate than that in the {\em FUSE} range.\\

In conclusion, we adopt atmospheric parameters for GD\,394 of $T_{\mathrm{eff}}=35700\pm1500$\,K, $\log g=8.05\pm0.1$. Our result is consistent with published fits to FUV data and we find, as in previous studies, disagreement between results obtained via fits to the Lyman and Balmer lines.

\begin{figure*}
    \centering
    \includegraphics[width=15.0cm]{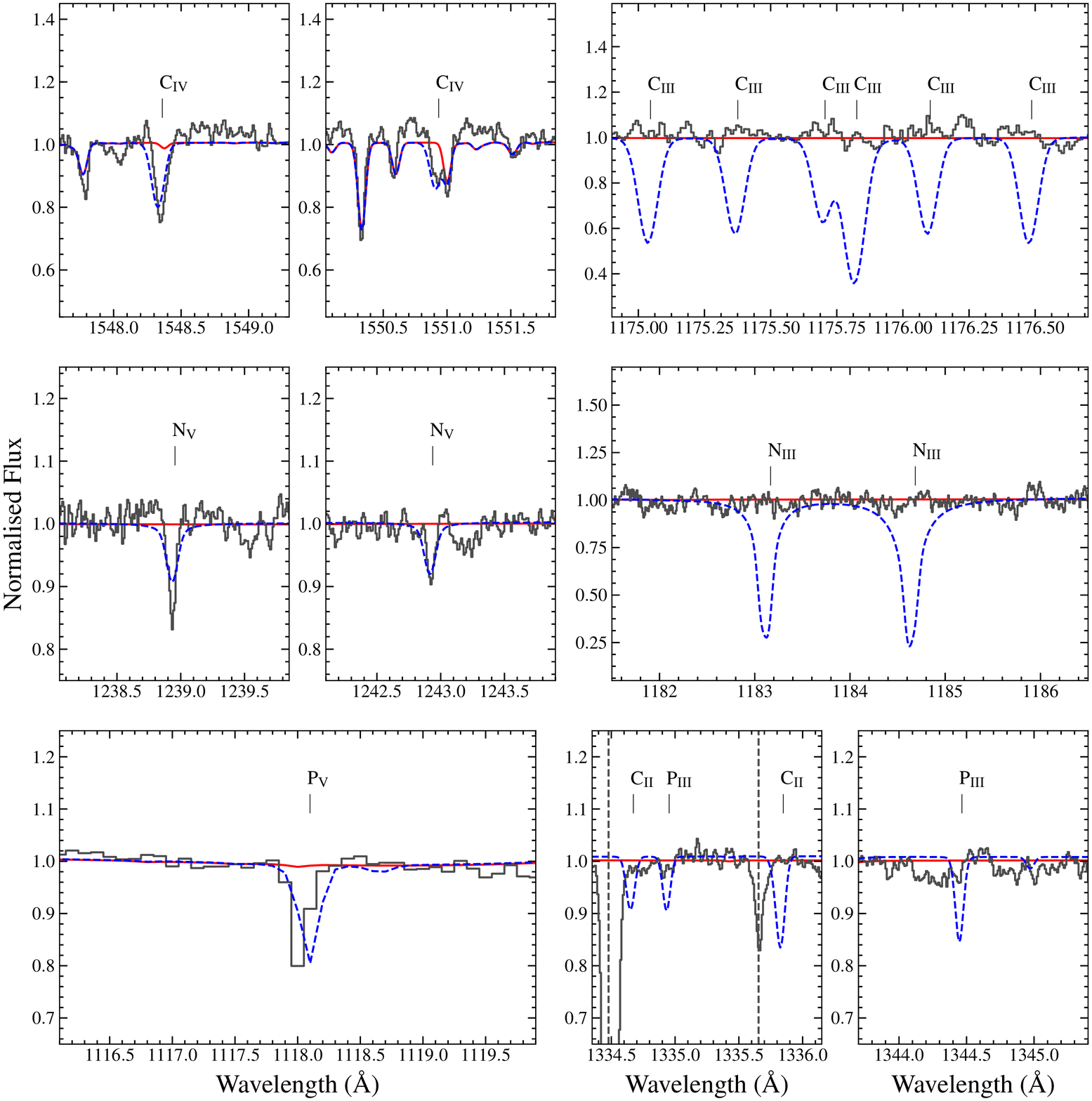}
    \caption{High-ionisation level lines of C, N and P in the STIS and {\em FUSE} spectra (left). The red lines show a model fit disregarding the high-ionisation lines and the blue dashed line shows a fit to just the high-ionisation lines. The abundances measured from the high-ionisation lines predict multiple strong low-ionisation level lines of the same elements, none of which are detected (right).} \protect\label{fig:high_ex}
\end{figure*}

\subsection{Metal abundances, diffusion, radiative levitation and the origin of the metal lines}
There are a plethora of silicon lines visible in the STIS FUV spectrum, including the three ionization  states, \ion{Si}{ii}, \ion{Si}{iii} and \ion{Si}{iv}. Fitting each ion separately, we find a consistent fit to all three ionization states, with the exception of \ion{Si}{iii} lines with excitation energies $\gtrsim 7$\,eV (Figure \ref{fig:gd394_si_levels}). Similarly, the Fe abundances measured from \ion{Fe}{iii} and \ion{Fe}{iv} lines are also in agreement. We confirm previous detections of \ion{Al}{iii} \citep[][Figure \ref{fig:gd394_alfe}]{holbergetal98-1, dupuisetal00-1,chayeretal00-1}, along with a number of lines that we tentatively identify as \ion{Ni}{iii}, but as they all coincide with strong Fe lines we treat the Ni abundance measured as an upper limit. Abundances for all detected metals are given in Table \ref{tab:abds}. The STIS and {\em FUSE} spectra also contain strong \ion{C}{iv}, \ion{N}{v} and \ion{P}{v} lines which we discuss in Section \ref{sec:high_ex}. 

The optical HIRES spectra contain multiple photospheric \ion{Si}{iii} lines with high excitation energies, from which we measured a Si abundance of $-5.10\pm0.2$. This is clearly incompatible with the abundances obtained from the low-excitation \ion{Si}{ii}, \ion{Si}{iii} and \ion{Si}{iv} lines in the STIS FUV spectrum, but agrees well with measurements of the high excitation \ion{Si}{iii} FUV lines (Figure \ref{fig:gd394_si_levels}). It is therefore unlikely to be due to genuine variation of the accretion rate between the observations, especially given that there is no change in line strength between the various STIS and HIRES spectra. The discrepancies in Si abundances were also reported by \citet{dupuisetal00-1}. 
Several other metal-polluted white dwarfs show similar discrepancies between Si measurements from optical and ultraviolet data \citep{gaensickeetal12-1, xuetal17-1}, but as these authors do not report comparisons between different \ion{Si}{iii} lines within the same spectrum, variation between epochs cannot be completely ruled out for the stars discussed in these papers.

Assuming that GD\,394 has a ``standard'' DAZ atmosphere, we calculated the strength of radiative levitation of Si according to the procedure described in \citet{koesteretal14-1} and adopting the atmospheric parameters and Si abundance from the best fit to the STIS spectrum. We find the maximum abundance of Si that can be supported to be $\log (\mathrm{Si/H}) \leq -6.1$, so under the standard accretion-diffusion equilibrium scenario with radiative levitation accounted for GD\,394 must be accreting Si at $\dot M(\mathrm{Si})=1.0\times10^6\mathrm{g\,s^{-1}}$, a low-to-medium rate compared to the bulk of the DAZ population \citep[see figure 8 in][]{koesteretal14-1}. However it is clear that GD\,394 is far from a typical DAZ white dwarf given the discrepancy between ultraviolet and optical fits, EUV variation and the anomalous high-excitation lines discussed below, so these calculations have to be considered with some caution. Radiative levitation for Al and Fe was not treated in \citet{koesteretal14-1}, although \citet{chayeretal95-2} find the radiative support to be low.

Without precise accretion fluxes speculation on the origin of the debris is limited, so we only note that the raw values of $\log (\mathrm{Al/Si}) = - 1.11$ and $\log (\mathrm{Fe/Si}) = -0.03$ are close enough to those of the bulk Earth \citep[-0.988 and -0.002 respectively,][]{mcdonough00-1} and the carbon content is sufficiently low that there is no reason to doubt that GD\,394 is accreting rocky debris from a remnant planetary system \citep{jura06-1,gaensickeetal12-1}.

\subsection{High-ionisation lines}\label{sec:high_ex}

\citet{chayeretal00-1} identified \ion{P}{v}\,1117.977\,\AA\ and 1128.008\,\AA\ absorption lines in the {\em FUSE} spectrum. The latter line is blended with \ion{Fe}{iii} and \ion{Si}{iv} transitions, so we assume that their P abundance is based on the 1117\,\AA\ line\footnote{We note that \citet{chayeretal00-1} give the wrong wavelength for this line, 1122\AA, in their tables 2 and 3.}. For our best fit model to \ion{P}{v}\,1117.977\AA\ we obtain $\log(\mathrm{P/H}) = -7.5\pm0.2$, in agreement with \citet{chayeretal00-1}. We also detect \ion{C}{iv}\,1548.202\,\AA\ and 1550.774\,\AA\ lines at $\log(\mathrm{C/H}) \approx -7.5$, and \ion{N}{v}\,1238.821\,\AA\ and 1242.804\,\AA\ lines at $\log(\mathrm{N/H}) \approx -3.7$ (Figure \ref{fig:high_ex}, left) in the STIS spectrum. All of these lines are at the photospheric velocity (Table \ref{tab:gd394_cs_lines}), with the possible exception of \ion{C}{iv} for which the best fit gives a redshift of $\approx 2\mathrm{\,km\,s}^{-1}$ relative to adjacent \ion{Fe}{iii} lines, but this is within the uncertainty of the spectral resolution. Unless this is an unlikely coincidence, these are therefore either photospheric features, or are being produced in a layer just above the photosphere, close enough such that there is no detectable difference in gravitational redshift. We do not detect secondary, clearly circumstellar lines (i.e. at different velocities to the photospheric lines) such as those detected at multiple hot white dwarfs by \citet{dickinsonetal12-2}.

However, no \ion{C}{ii}, \ion{C}{iii}, \ion{N}{iii} or \ion{P}{iii} lines are detected in the STIS spectrum, all of which are predicted to be strong when adopting photospheric abundances based on the high-ionisation lines (Figure \ref{fig:high_ex}, right). This non-detection places upper limits on the abundances of C, N and P of $\log(\mathrm{X/H}) \leq-8.00$, clearly incompatible with the measurements from the high-ionisation lines (Table \ref{tab:abds}). We conclude that the material producing the high-ionisation level lines must originate in a hot layer close to, but outside of the white dwarf photosphere. 

As mentioned above the higher excitation \ion{Si}{iii} lines are also too strong for our adopted parameters. However, as similar discrepancies between Si excitation states have been detected in other white dwarfs without similar anomalous C, N and P lines it is unclear if the explanation for the Si mismatch is the same as for the other elements.

\begin{table}
\centering
\caption{Abundances of metals in the atmosphere of GD\,394. The third column shows the photospheric abundances required to produce the high-excitation/ionisation absorption lines, which are clearly ruled out by upper limits from the lower energy absorption lines.}
\begin{tabular}{llll}\\
\hline
\multicolumn{2}{c}{Element} & \multicolumn{2}{c}{Abundance ($\log[\mathrm{X/H}]$)}\\
& & Photosphere lines & High-ex/ion lines \\
\hline
6 & C &  $\leq-8.00 $  & $-7.50$ \\
7 & N &  $\leq-8.00 $  & $-3.7$ \\
8 & O &  $\leq-5.00$ & \\
13 & Al  & $-7.07\pm0.20$ & \\
14 & Si  & $-5.96\pm0.10$ & $-5.10\pm0.20$ \\
15 & P  & $ \leq-8.00$ & $-7.5\pm0.2$\\
26 & Fe & $-5.93\pm0.20$ & \\
28 & Ni & $ \leq-7.80 $ & \\
\hline
\end{tabular}
\label{tab:abds}
\end{table}

\section{Non-detection of gaseous emission}\label{sec:gd394_gas}
\ion{Ca}{ii}\,8600\,\AA\ emission from a gaseous component to a debris disc has been confirmed at seven metal-polluted white dwarfs to date \citep{gaensickeetal06-3,gaensickeetal07-1,gaensickeetal08-1,dufouretal12-1,farihietal12-1,melisetal12-1,wilsonetal14-1}. The emission takes a distinct double-peaked morphology induced by the Keplerian orbital motion of the disc material \citep{horne+marsh86-1}. \citet{burleighetal11-1} observed GD\,394 as part of a search for gaseous emission at hot white dwarfs, returning no detections. As gaseous discs can form on $\approx$~year-long timescales \citep{wilsonetal14-1}, it is worth noting that our 2016 WHT/ISIS observation also failed to detect emission. Non-detection of gaseous emission is unsurprising, as in all known cases it is associated with the presence of an infrared excess from dusty debris \citep{brinkworthetal12-1},  which was ruled out at GD\,394 by \citet{mullallyetal07-1}. Dust at GD\,394 will sublimate at radii greater than the Roche radius, preventing the formation of a compact dusty debris disc \citep{vonhippeletal07-1}.

\section{Discussion}\label{sec:gd394_discussion}
GD\,394 is not variable at any of the wavelengths and on any of the timescales explored here, contrasting strongly with the large-amplitude EUV variability observed by \citet{dupuisetal00-1}. 

A potential explanation is that the accretion spot hypothesis was correct, but that the spot has dispersed since the {\em EUVE} observations and GD\,394 is now accreting uniformly over its surface. The low accretion rate and lack of \ion{Ca}{ii} emission lines suggest that the circumstellar environment may be relatively inactive, with the short diffusion timescales removing any evidence of higher activity in the past. However, this explanation conflicts with the perfect match between the metal absorption lines in the GHRS and STIS observations. The GHRS observations were obtained less than a year before the first {\em EUVE} observations, so it is unlikely that the EUV variation was not present at that epoch, especially as it was detected in all four {\em EUVE} observations over the following four years. Any long-term variation in the accretion rate or surface distribution of metals should likely have resulted in noticeable differences between the STIS and GHRS spectra.  

If the EUV variation was still present during all of the observations presented here, then an accretion spot can be ruled out as the cause. An alternative possibility is that GD\,394 hosts a planet on a 1.15\,day orbit. Multiple hot Jupiter planets have been observed with ultraviolet transits 5--10\,per\,cent deeper than in the optical \citep{haswelletal12-1}, interpreted as the transit of a cloud of evaporating material around the planet. Given the small size of GD\,394, it is conceivable that an orbiting planet may not transit itself but be surrounded by a hydrogen cloud that clips the white dwarf, causing the EUV variations. The Ly\,$\alpha$ absorption seen in main sequence examples \citep{vidal-madjaretal03-1} is masked by the deep, wide photospheric and interstellar  absorption at GD\,394. Assuming that the 1.15\,d signal detected in the EUV is the orbital period of a planet, then the equilibrium surface temperature of the planet will be $\approx 1300$\,K, enough to induce atmospheric evaporation \citep{tripathietal15-1}. The material lost by this planet would also accrete onto GD\,394, providing the reservoir for the photospheric metal pollution (see \citealt{farihietal17-1} for an example of a white dwarf with both a low-mass companion and planetary debris). Testing this hypothesis requires high--precision radial velocity measurements, although this will be challenging due to the paucity of photospheric absorption lines at optical wavelengths. Alternatively new X-ray/EUV observations could probe for the distinctive asymmetric transit produced by evaporating planets, which may not have been resolved in the {\em EUVE} observations.

Is GD\,394 unique? BOKS\,53856 is a faint variable white dwarf, which is often discussed in conjunction with GD\,394. \citep{holberg+howel11-1} observed this star early in the {\em Kepler} mission, identifying it as a moderate temperature DA white dwarf, somewhat cooler than GD\,394, having a non-sinusoidal light curve with a 5\,per\,cent minimum to maximum variation and a period of 0.2557\,d which persisted over the six months of observation. \citet{holberg+howel11-1} suggested that the light curve could be explained by the rotation of BOKS\,53856, where a frozen-in photospheric magnetic field produced a localized ``spot'', in analogy with the explanation of the EUV variations at GD\,394 offered by \citet{dupuisetal00-1}. For both stars, it was assumed that the spots represent magnetically confined regions of higher opacity due to accreted metals.

Recently \citet{hoardetal18-1} conducted an extensive campaign of space-based and ground-based observation of BOKS\,53856, extending the earlier {\em Kepler} observations to cover a 4\,yr time span, along with a corresponding ultraviolet lightcurve covering the entire spin period from TIME-TAG ultraviolet spectroscopy obtained with the Cosmic Origins Spectrograph (COS) onboard {\em HST}. These data yielded a very precise spin period and established that the optical pulsations remained coherent and unchanged over this period.  Using a simple model consisting of two-phase temperature distribution, they were able to produce brightness maps of the BOKS\,53856 photosphere for both the ultraviolet and optical data that to first order show a similar distribution of spots.

In contrast to the metal-rich STIS spectra of GD\,394 presented here, \citep{hoardetal18-1} identify no photospheric features in the COS spectra of BOKS\,53586 beyond hydrogen. Thus, although both GD\,394 and BOKS\,53856 exhibit flux variations that are difficult to explain without invoking magnetic fields and accretion, it is GD\,394 that has the expected metal rich ultraviolet spectrum but no presently detectable photometric or spectroscopic variations, whilst BOKS\,53856 shows stable, persistent photometric variations but no trace of any expected metals. It may require sensitive spectropolarimetry over the rotational periods of both stars to definitively detect any putative magnetic fields.

The high ionisation lines observed in the STIS and {\em FUSE} spectra of GD\,394 may be produced in a high-temperature accretion flow similar to that observed in cataclysmic variables \citep{patterson+raymond85-1}, but that would still lead to accretion of C, N and P into the photosphere. Radiative levitation is potentially strong enough to expel all C from the photosphere but the predicted support for N is far too low \citep{chayeretal95-2}. \citet{lallementetal11-1} detected \ion{C}{iv} lines without corresponding \ion{C}{iii} lines at two white dwarfs with similar $T_{\mathrm{eff}}$ to GD\,394. In one, WD\,1942+499, they also detected \ion{P}{v} and \ion{O}{vi} lines, again without predicted lower ionisation level lines. In contrast, the $T_{\mathrm{eff}}=28000-30000$\,K white dwarf component of the dwarf nova U\,Geminorum has \ion{N}{v} lines at photospheric wavelengths, which require $T_{\mathrm{eff}}\approx80000$\,K \citep{sionetal98-1,long+gilliland99-1}, but does have low-excitation N lines that require a super-Solar N abundance to fit \citep{longetal06-1}. If radiative levitation is expelling both N and C at GD\,394 then we would expect the same to happen white dwarfs with similar $T_{\mathrm{eff}}$; this is clearly not the case.

\section{Conclusion}\label{sec:gd394_conclusion}
We obtained multi-epoch, multi-wavelength observations of GD\,394 to test the accretion spot hypothesis put forward by \citet{dupuisetal00-1} to explain the large amplitude flux variations detected in the EUV. We find no evidence for any change in photospheric metal abundances over the 1.15\,d period of the EUV variation, nor on the decades-long timescales covered by {\em HST} spectroscopy. No photometric variability is observed at any waveband  beyond the EUV. The EUV variation may have either stopped, although the  agreement between near-contemporaneous spectra and more recent observations disfavours this explanation, or is being caused by some phenomena other than a spot, such as an otherwise undetected evaporating planet. Distinguishing between these scenarios will require new observations in the EUV.

Beyond a search for variation, our observations show GD\,394 to be a highly unusual white dwarf. As with previous studies for multiple hot white dwarfs, we cannot obtain consistent atmospheric parameters between fits to the optical and ultraviolet hydrogen lines. The analysis of the observed metal lines also leads to contradictory results, especially the presence of high ionisation level lines of C, N, and P sharing the photospheric velocity of the white dwarf, but being strictly incompatible with the non-detection of the corresponding low ionisation level lines.

\section*{Acknowledgments}
DJW, BTG, PC and MAH have received funding from the European Research Council under the European Union's Seventh Framework Programme (FP/2007-2013) / ERC Grant Agreement n. 320964 (WDTracer). OT was supported by a Leverhulme Trust Research Project Grant.

This paper is based on observations made with the NASA/ESA {\em Hubble Space Telescope}, obtained at the Space Telescope Science Institute, which is operated by the Association of Universities for Research in Astronomy, Inc., under NASA contract NAS 5-26555. These observations are associated with program ID\,13719.  Support for KSL's and JH's effort on program ID 13719 was provided by NASA through a grant from the Space Telescope Science Institute.

     The William Herschel Telescope is operated on the island of La Palma by the Isaac Newton Group in the Spanish Observatorio del Roque de los Muchachos of the Instituto de Astrof\'{\i}sica de Canarias. Data for this paper have been obtained under the International Time Programme of the CCI (International Scientific Committee of the Observatorios de Canarias of the IAC).

Some of the data presented herein were obtained at the W.M. Keck Observatory, which is operated as a scientific partnership among the California Institute of Technology, the University of California and the National Aeronautics and Space Administration. The Observatory was made possible by the generous financial support of the W.M. Keck Foundation. The authors wish to recognize and acknowledge the very significant cultural role and reverence that the summit of Mauna Kea has always had within the indigenous Hawaiian community.  We are most fortunate to have the opportunity to conduct observations from this mountain.

\defcitealias{astropy13-1}{Astropy Collaboration, 2013}
This research made use of Astropy, a community-developed core Python package for Astronomy \citepalias{astropy13-1}.




\bibliographystyle{mnras}
\bibliography{aamnem99,aabib} 


\clearpage
\appendix
\section{Log of spectroscopic observations}
\begin{table*}
\centering
\caption{Summary of the spectroscopic observations of GD\,394. For the STIS FUV observations we give the phase position relative to the 1.15\,d period, with the start of the first observation defining phase zero. } 
\begin{tabular}{lccccc}\\
\hline
Date & Telescope/Instrument & Start Time (UT) & Total Exposure Time (s) & Phase & Wavelength range (\AA) \\
\hline
2016 August 13 & WHT/ISIS & $23:24:00$ & 1800 & - & 3056--5409, 5772--9088\\ 

2015 November 15 & Keck/HIRES & $04:38:13$ & 900 & - & 3125--5997 \\

2015 August 25 & {\em HST}/STIS & $16:43:59$ & 2595 & 4.3 & 1160--1710\\ 
2015 August 24 & {\em HST}/STIS & $20:01:05$ & 2595 & 3.5 & 1160--1710\\
2015 August 24 & {\em HST}/STIS & $09:02:39$ & 2595 & 3.1 & 1160--1710\\
2015 August 23 & {\em HST}/STIS & $12:20:34$ & 2595 & 2.4 & 1160--1710\\
2015 August 22 & {\em HST}/STIS & $15:38:36$ & 2595 & 1.6 & 1160--1710\\
2015 August 21 & {\em HST}/STIS & $18:56:37$ & 2595 & 0.9 & 1160--1710\\
2015 August 21 & {\em HST}/STIS & $15:45:25$ & 2595 & 0.8 & 1160--1710\\
2015 August 20 & {\em HST}/STIS & $19:05:13$ & 2595 & 0.0 & 1160--1710\\

2013 December 23 & {\em HST}/STIS & $10:58:53$ & 1500 & - & 2577--2835 \\

2009 May 23 & Keck/HIRES & $14:40:12$ & 1800 & - & 3125--5997, 4457--7655\\

2007 August 06 & WHT/ISIS & $00:32:00$ & 2882 & - & 4520--5262, 8260--9014 \\ 

2002 October 27 & {\em FUSE} & $16:44:22$ & 4403 & - &	905--1181\\
2002 September 04 & {\em FUSE} & $05:18:54$ & 7957 & - & 905--1181\\
2000 September 09 & {\em FUSE} & $19:51:31$ & 4155 & - & 905--1181\\
2000 June 21 & {\em FUSE} & $14:03:44$ & 3327 & - & 905--1181\\
2000 June 20 & {\em FUSE} & $18:04:49$ & 28310 & - & 905--1181\\
1999 October 11 & {\em FUSE} & $04:29:36$ & 3662 & - & 905--1181\\
1999 October 11 & {\em FUSE} & $10:33:29$ & 5652 & - &	905--1181\\
1999 October 13 & {\em FUSE} & $11:00:21$ & 4688 & - & 905--1120\\

1992 June 18 & {\em HST}/GHRS & $07:30:19$ & 653 & - & 1196--1232 \\
1992 June 18 & {\em HST}/GHRS & $08:32:47$ & 490 & - & 1290--1325\\
1992 June 18 & {\em HST}/GHRS & $08:44:29$ & 1088 & - & 1383--1418\\ 
\hline
\end{tabular}
\label{tab:fuv_obs}
\end{table*}

\clearpage

\section{Line lists}\label{sec:line_lists}
\begin{table*}
\centering
\caption{Observed photospheric metal absorption lines detected in the FUSE, STIS and HIRES spectra. The rest wavelengths for lines detected in HIRES spectra use the air value, all others use the vacuum value. Each line was fitted with a Gaussian curve using {\sc Astropy} least-squares fitting routines, and the equivalent width measured according to the prescriptions found in \citet{vollmann+eversberg06-1}. Lines where an accurate $\Delta$v could not be obtained are marked ``$--$''.  *Blended with \ion{C}{IV}\,1550.77\,\AA. }
\label{tab:gd394_ps_lines}
\begin{tabular}{lcccc}
\hline
Line & Rest $\lambda$\,(\AA) & Observed $\lambda$\,(\AA) & $\Delta$v\,(kms$^{-1}$) & Equivalent Width\,(m\AA)\\ \hline
\multicolumn{5}{l}{{\em FUSE}} \\ 
\ion{Fe}{iii} & 983.88 & 983.925 & $--$ & $25.0\pm3.7$\\ 
\ion{Fe}{iii} & 985.0 & 985.889 & $--$ & $16.0\pm4.4$\\ 
\ion{Fe}{iii} & 986.0 & 986.651 & $--$ & $3.5\pm3.5$\\ 
\ion{N}{iii}? & 989.7 & 989.784 & $25.4\pm17.1$ & $36.0\pm2.7$\\ 
\ion{N}{iii} & 991.5 & 991.899 & $--$ & $35.0\pm3.1$\\ 
\ion{Si}{iii} & 993.52 & 993.585 & $19.6\pm12.6$ & $46.0\pm2.4$\\ 
\ion{Si}{iii} & 994.79 & 994.845 & $16.7\pm6.68$ & $81.0\pm3.1$\\ 
\ion{Si}{iii} & 997.39 & 997.468 & $23.4\pm3.6$ & $87.0\pm3.5$\\ 
\ion{Fe}{iii} & 1017.25 & 1017.3 & $15.8\pm11.8$ & $12.0\pm2.7$\\ 
\ion{Fe}{iii} & 1018.29 & 1018.32 & $9.82\pm29.1$ & $-0.4\pm2.7$\\ 
\ion{Si}{iii} & 1108.37 & 1108.41 & $10.9\pm12.9$ & $76.0\pm1.9$\\ 
\ion{Si}{iii} & 1109.97 & 1110.01 & $10.8\pm5.87$ & $110.0\pm2.1$\\ 
\ion{Si}{iii} & 1113.23 & 1113.27 & $10.0\pm0.798$ & $160.0\pm2.7$\\ 
\ion{Si}{iii}/\ion{Si}{iv} & 1122.49 & 1122.53 & $9.66\pm1.19$ & $120.0\pm2.7$\\ 
\ion{Si}{iv} & 1128.34 & 1128.38 & $11.3\pm4.52$ & $150.0\pm3.1$\\ 
\ion{Si}{iii} & 1144.31 & 1144.43 & $30.8\pm32.1$ & $18.0\pm1.1$\\ 
\multicolumn{5}{l}{{\em STIS}} \\ 
\ion{Si}{ii} & 1194.496 & 1194.59 & $24.3\pm14.0$ & $20.0\pm17.0$\\ 
\ion{Si}{iii} & 1206.5 & 1206.62 & $28.9\pm1.08$ & $260.0\pm45.0$\\ 
\ion{Si}{iii} & 1207.517 & 1207.62 & $25.5\pm8.08$ & $34.0\pm8.1$\\ 
\ion{Fe}{iii}? & 1210.4 & 1210.56 & $40.9\pm1.53$ & $20.0\pm13.0$\\ 
\ion{Si}{iii} & 1235.43 & 1235.58 & $35.3\pm1.53$ & $26.0\pm7.9$\\ 
\ion{Si}{iii} & 1235.43 & 1235.58 & $35.3\pm1.53$ & $26.0\pm7.9$\\ 
\ion{Si}{iii} & 1238.8 & 1238.92 & $30.1\pm2.65$ & $10.0\pm8.6$\\ 
\ion{Fe}{ii}? & 1242.8 & 1242.91 & $27.3\pm5.28$ & $7.2\pm13.0$\\ 
\ion{Si}{ii} & 1250.43 & 1250.54 & $27.4\pm3.36$ & $10.0\pm8.8$\\ 
\ion{Si}{ii} & 1264.738 & 1264.85 & $26.6\pm1.07$ & $25.0\pm6.7$\\ 
\ion{Si}{iii} & 1280.35 & 1280.47 & $28.2\pm2.19$ & $13.0\pm9.0$\\ 
\ion{Si}{iii} & 1294.545 & 1294.67 & $28.9\pm0.38$ & $81.0\pm8.3$\\ 
\ion{Si}{iii} & 1296.726 & 1296.87 & $32.4\pm0.68$ & $79.0\pm7.3$\\ 
\ion{Si}{iii} & 1298.892 & 1299.06 & $38.3\pm0.587$ & $150.0\pm12.0$\\ 
\ion{Si}{iii} & 1301.149 & 1301.27 & $27.6\pm3.2$ & $72.0\pm6.7$\\ 
\ion{Si}{iii} & 1303.323 & 1303.44 & $26.5\pm3.37$ & $77.0\pm6.2$\\ 
\ion{Si}{ii} & 1305.2 & 1305.33 & $30.1\pm59.3$ & $3.7\pm2.6$\\ 
\ion{Si}{ii} & 1309.27 & 1309.39 & $27.8\pm8.38$ & $4.5\pm6.7$\\ 
\ion{Si}{iii} & 1312.591 & 1312.71 & $28.0\pm0.487$ & $44.0\pm6.6$\\ 
\ion{Si}{iii} & 1341.458 & 1341.59 & $29.2\pm2.11$ & $38.0\pm7.2$\\ 
\ion{Ni}{iii} & 1342.1 & 1342.51 & $--$ & $35.0\pm8.0$\\ 
\ion{Si}{iii} & 1343.409 & 1343.52 & $24.4\pm4.56$ & $22.0\pm6.7$\\ 
\ion{Si}{iii} & 1361.596 & 1361.72 & $26.3\pm2.12$ & $24.0\pm6.8$\\ 
\ion{Si}{iii} & 1362.37 & 1362.49 & $25.9\pm6.99$ & $11.0\pm6.6$\\ 
\ion{Si}{iii} & 1363.459 & 1363.58 & $27.2\pm4.42$ & $21.0\pm6.5$\\ 
\ion{Si}{iii} & 1365.253 & 1365.39 & $29.2\pm2.14$ & $34.0\pm7.4$\\ 
\ion{Si}{iii} & 1367.027 & 1367.17 & $30.6\pm5.0$ & $16.0\pm5.8$\\ 
\ion{Si}{iii} & 1369.44 & 1369.57 & $27.8\pm4.29$ & $13.0\pm12.0$\\ 
\ion{Si}{iii} & 1373.03 & 1373.15 & $26.6\pm3.25$ & $6.7\pm6.1$\\ 
\ion{Al}{iii} & 1379.67 & 1379.79 & $25.6\pm2.56$ & $8.2\pm11.0$\\ 
\ion{Al}{iii} & 1384.132 & 1384.27 & $29.4\pm1.32$ & $21.0\pm8.0$\\ 
\ion{Si}{iii} & 1387.99 & 1388.13 & $29.5\pm0.0$ & $11.0\pm19.0$\\ 
\ion{Si}{iv} & 1393.755 & 1393.88 & $26.9\pm0.576$ & $480.0\pm49.0$\\ 

\hline
\end{tabular}
\end{table*}

\begin{table*}
\centering
\contcaption{}
\label{tab:continued}
\begin{tabular}{lcccc}
\hline
Line & Rest $\lambda$\,(\AA) & Observed $\lambda$\,(\AA) & $\Delta$v\,(kms$^{-1}$) & Equivalent Width\,(m\AA)\\ \hline
\ion{Si}{iv} & 1402.77 & 1402.89 & $26.4\pm0.38$ & $310.0\pm34.0$\\ 
\ion{Ni}{iii} & 1433.6 & 1433.8 & $41.7\pm2.29$ & $14.0\pm10.0$\\ 
\ion{Si}{iii} & 1435.772 & 1435.91 & $28.7\pm1.51$ & $32.0\pm9.1$\\ 
\ion{Ni}{iii} & 1436.7 & 1436.86 & $32.9\pm10.6$ & $11.0\pm11.0$\\ 
Fe/Ni? & 1457.2 & 1457.42 & $44.4\pm4.92$ & $5.2\pm7.8$\\ 
\ion{Fe}{iii} & 1465.7 & 1465.89 & $38.9\pm13.0$ & $13.0\pm21.0$\\
\ion{Ni}{iii} & 1467.7 & 1467.89 & $37.9\pm0.0$ & $4.8\pm5.2$\\ 
\ion{Fe}{iii} & 1467.7 & 1467.89 & $37.9\pm7.47$ & $4.9\pm5.3$\\ 
\ion{Ni}{iii} & 1469.8 & 1470.0 & $40.7\pm6.31$ & $11.0\pm16.0$\\ 
\ion{Fe}{iii} & 1469.8 & 1470.0 & $40.7\pm6.49$ & $11.0\pm16.0$\\ 
\ion{Fe}{iii} & 1481.1 & 1481.34 & $47.9\pm10.4$ & $9.5\pm17.0$\\ 
\ion{Si}{iii} & 1500.241 & 1500.38 & $27.4\pm3.3$ & $53.0\pm9.8$\\

\ion{Si}{iii} & 1501.15 & 1501.32 & $33.9\pm3.0$ & $59.0\pm12.0$\\ 
\ion{Si}{iii} & 1501.197 & 1501.32 & $24.3\pm3.12$ & $61.0\pm14.0$\\ 
\ion{Si}{iii} & 1501.78 & 1502.0 & $43.7\pm4.13$ & $49.0\pm11.0$\\ 
\ion{Si}{iii} & 1501.827 & 1501.99 & $33.0\pm3.88$ & $50.0\pm14.0$\\ 
\ion{Fe}{iii} & 1504.0 & 1504.16 & $31.9\pm16.1$ & $6.3\pm11.0$\\ 
\ion{Fe}{iii} & 1505.1 & 1505.29 & $37.0\pm5.42$ & $13.0\pm9.4$\\ 
\ion{Si}{iii} & 1506.06 & 1506.2 & $27.8\pm4.1$ & $11.0\pm8.4$\\ 
\ion{Fe}{iii} & 1511.6 & 1511.76 & $30.9\pm6.8$ & $6.9\pm13.0$\\ 
\ion{Fe}{iv}? & 1523.9 & 1524.07 & $33.6\pm5.56$ & $4.7\pm4.3$\\ 
\ion{Fe}{iv} & 1523.923 & 1524.07 & $29.1\pm5.59$ & $4.5\pm4.1$\\ 
\ion{Fe}{iii} & 1524.5 & 1524.65 & $30.3\pm10.9$ & $3.6\pm4.0$\\ 
\ion{Fe}{iii} & 1524.6 & 1524.78 & $36.3\pm0.0$ & $6.2\pm9.0$\\ 
\ion{Fe}{iii} & 1525.036 & 1525.16 & $24.7\pm50.0$ & $6.3\pm8.1$\\ 
\ion{Fe}{iii} & 1525.798 & 1525.94 & $27.7\pm25.0$ & $6.4\pm3.7$\\ 
\ion{Fe}{iii}? & 1526.5 & 1526.67 & $33.3\pm0.767$ & $75.0\pm13.0$\\ 
\ion{Fe}{iii} & 1527.141 & 1527.28 & $27.9\pm30.3$ & $7.5\pm6.8$\\ 
\ion{Fe}{iii} & 1531.64 & 1531.78 & $27.1\pm6.8$ & $7.7\pm8.3$\\ 
\ion{Fe}{iii} & 1531.864 & 1531.99 & $24.5\pm8.34$ & $7.2\pm5.1$\\ 
\ion{Fe}{iv} & 1532.63 & 1532.77 & $27.6\pm16.8$ & $4.0\pm3.8$\\ 
\ion{Fe}{iv} & 1536.577 & 1536.72 & $28.8\pm8.04$ & $8.8\pm7.8$\\ 
\ion{Fe}{iii} & 1538.629 & 1538.77 & $26.5\pm4.15$ & $14.0\pm8.3$\\ 
\ion{Fe}{iii} & 1539.123 & 1539.26 & $26.6\pm4.53$ & $10.0\pm6.4$\\ 
\ion{Fe}{iii} & 1539.473 & 1539.61 & $27.3\pm11.5$ & $5.4\pm6.8$\\ 
\ion{Fe}{iii} & 1540.164 & 1540.3 & $26.2\pm13.2$ & $2.7\pm6.5$\\ 
\ion{Fe}{iii} & 1541.831 & 1541.96 & $24.7\pm10.7$ & $5.8\pm7.5$\\ 
\ion{Fe}{iv} & 1542.155 & 1542.3 & $27.4\pm4.89$ & $11.0\pm8.8$\\ 
\ion{Fe}{iv} & 1542.698 & 1542.84 & $28.2\pm2.95$ & $16.0\pm8.0$\\ 
\ion{Fe}{iii} & 1543.623 & 1543.78 & $29.7\pm6.75$ & $11.0\pm9.9$\\ 
\ion{Fe}{iii} & 1547.637 & 1547.78 & $28.0\pm6.98$ & $11.0\pm9.1$\\ 
\ion{Fe}{iii} & 1550.193 & 1550.33 & $26.0\pm3.29$ & $18.0\pm8.0$\\ 
\ion{Fe}{iii}* & 1550.862 & - & - & -\\ 
\ion{Fe}{iii} & 1552.065 & 1552.19 & $25.1\pm9.32$ & $8.3\pm8.1$\\ 
\ion{Fe}{iv} & 1552.349 & 1552.49 & $28.1\pm12.1$ & $4.1\pm4.9$\\ 
\ion{Fe}{iv} & 1552.705 & 1552.84 & $26.0\pm12.4$ & $5.3\pm8.5$\\ 
\ion{Fe}{iv} & 1553.296 & 1553.44 & $27.9\pm8.7$ & $6.5\pm11.0$\\ 
\ion{Fe}{iii} & 1556.076 & 1556.21 & $26.3\pm8.62$ & $3.5\pm6.7$\\ 
\ion{Fe}{iii} & 1556.498 & 1556.62 & $24.0\pm4.58$ & $6.6\pm4.5$\\ 
\ion{Al}{iii} & 1605.766 & 1605.91 & $27.6\pm2.48$ & $27.0\pm16.0$\\ 
\ion{Fe}{iii}? & 1607.7 & 1607.87 & $30.8\pm5.13$ & $20.0\pm14.0$\\ 
\ion{Al}{iii} & 1611.873 & 1611.99 & $22.6\pm1.52$ & $64.0\pm30.0$\\ 
\ion{Ni}{iii} & 1614.0 & 1614.18 & $32.9\pm11.8$ & $9.5\pm17.0$\\ 
\multicolumn{5}{l}{{\em HIRES}} \\ 
\ion{Si}{iii} & 3791.439 & 3791.76 & $25.4\pm0.98$ & $5.8\pm2.2$\\ 
\ion{Si}{iii} & 3796.124 & 3796.47 & $27.3\pm0.42$ & $18.0\pm3.3$\\ 
\ion{Si}{iii} & 3806.525 & 3806.89 & $28.8\pm0.61$ & $28.0\pm4.6$\\ 
\ion{Si}{iii} & 3924.468 & 3924.85 & $29.2\pm1.53$ & $6.6\pm4.4$\\ 
\ion{Si}{iii} & 4552.622 & 4553.04 & $27.5\pm0.40$ & $38.0\pm4.3$\\ 
\ion{Si}{iii} & 4567.84 & 4568.27 & $28.2\pm0.24$ & $32.0\pm3.2$\\ 
\ion{Si}{iii} & 4574.757 & 4575.19 & $28.4\pm0.62$ & $18.0\pm3.3$\\ 
\ion{Si}{iii} & 5739.734 & 5740.31 & $30.1\pm1.1$ & $20.0\pm1.8$\\ 
\hline

\end{tabular}
\end{table*}

\begin{table*}
\centering
\caption{High-ionisation absorption lines suspected to be circumstellar detected in the FUSE and STIS spectra.*Blended with \ion{Fe}{iii}\,1550.862\,\AA. }
\begin{tabular}{lcccc}
\hline
Line & Rest $\lambda$\,(\AA) & Observed $\lambda$\,(\AA) & $\Delta$v\,(kms$^{-1})$ & Equivalent Width\,(m\AA)\\ \hline
\multicolumn{5}{l}{{\em FUSE}} \\ 
\ion{P}{v} & 1117.977 & 1118.03 & $13.5\pm3.05$ & $33.0\pm1.9$\\ 
\multicolumn{5}{l}{{\em STIS}} \\ 
\ion{N}{v} & 1238.821 & 1238.93 & $26.3\pm2.85$ & $11.0\pm8.1$\\ 
\ion{N}{v} & 1242.804 & 1242.92 & $28.7\pm4.56$ & $6.7\pm8.5$\\ 
\ion{C}{iv} & 1548.187 & 1548.35 & $31.5\pm2.34$ & $19.0\pm12.0$\\ 
\ion{C}{iv}* & 1550.77 & - & - & - \\ 
\hline
\end{tabular}
\label{tab:gd394_cs_lines}
\end{table*}

\begin{table*}
\centering
\caption{ISM absorption lines detected in the FUSE, STIS and HIRES spectra.}
\begin{tabular}{lcccc}
\hline
Line & Rest $\lambda$\,(\AA) & Observed $\lambda$\,(\AA) & $\Delta$v\,(kms$^{-1})$ & Equivalent Width\,(m\AA)\\ \hline
\multicolumn{5}{l}{{\em FUSE}} \\ 
\ion{O}{i} & 988.773 & 988.66 & $-34.4\pm4.76$ & $94.0\pm3.8$\\ 
\ion{N}{ii}? & 1084.5 & 1083.93 & $--$ & $63.0\pm2.1$\\
\multicolumn{5}{l}{{\em STIS}} \\ 
\ion{Si}{ii} & 1190.4158 & 1190.38 & $-9.69\pm0.79$ & $68.0\pm13.0$\\ 
\ion{Si}{ii} & 1193.2897 & 1193.25 & $-9.07\pm0.953$ & $89.0\pm12.0$\\ 
\ion{N}{i} & 1199.5496 & 1199.51 & $-10.1\pm2.77$ & $74.0\pm10.0$\\ 
\ion{N}{i} & 1200.2233 & 1200.18 & $-10.1\pm3.12$ & $67.0\pm10.0$\\ 
\ion{N}{i} & 1200.7098 & 1200.67 & $-10.0\pm5.24$ & $51.0\pm10.0$\\ 
\ion{S}{ii} & 1253.805 & 1253.77 & $-9.48\pm3.7$ & $19.0\pm16.0$\\ 
\ion{S}{ii} & 1259.518 & 1259.47 & $-10.9\pm19.6$ & $12.0\pm5.9$\\ 
\ion{Si}{ii} & 1260.4221 & 1260.39 & $-8.62\pm0.949$ & $110.0\pm8.6$\\ 
\ion{O}{i} & 1302.168 & 1302.13 & $-8.77\pm2.29$ & $110.0\pm7.1$\\ 
\ion{Si}{ii} & 1304.3702 & 1304.33 & $-8.38\pm4.68$ & $39.0\pm6.3$\\ 
\ion{C}{ii} & 1334.532 & 1334.49 & $-9.3\pm0.397$ & $130.0\pm7.9$\\ 
\ion{C}{ii} & 1335.708 & 1335.65 & $-13.5\pm25.2$ & $13.0\pm8.8$\\ 
\ion{Si}{ii} & 1526.7066 & 1526.67 & $-7.18\pm0.821$ & $75.0\pm13.0$\\ 
\ion{Al}{ii} & 1670.7874 & 1670.74 & $-8.14\pm1.01$ & $54.0\pm20.0$\\ 
\ion{Fe}{ii} & 2586.65 & 2586.58 & $-7.55\pm4.01$ & $61.0\pm46.0$\\ 
\ion{Fe}{ii} & 2600.173 & 2600.1 & $-8.17\pm2.23$ & $120.0\pm49.0$\\ 
\ion{Mg}{ii} & 2796.352 & 2796.26 & $-10.1\pm1.54$ & $170.0\pm74.0$\\ 
\ion{Mg}{ii} & 2803.531 & 2803.44 & $-9.67\pm1.85$ & $180.0\pm67.0$\\
\multicolumn{5}{l}{{\em HIRES}} \\ 
\ion{Ca}{ii} & 3933.663 & 3933.53 & $-10.2\pm-0.18$ & $12.0\pm2.1$\\ 
\ion{Ca}{ii} & 3968.469 & 3968.32 & $-11.3\pm-0.53$ & $5.0\pm0.51$\\ 
\hline
\end{tabular}
\label{tab:gd394_ism_lines}
\end{table*}

\clearpage


\bsp	
\label{lastpage}
\end{document}